\documentclass[12pt,a4paper]{article}
\usepackage{epsfig}

\def\eq#1{{Eq.~(\ref{#1})}}
\newcommand{\Le}{\left(}
\newcommand{\Ra}{\right)}

\newcommand{\beq}{\begin{equation}}
\newcommand{\eeq}{\end{equation}}
\newcommand{\beqar}[1]{\begin{eqnarray}\label{#1}}
\newcommand{\eeqar}{\end{eqnarray}}
\newcommand{\g}{{\rm g}}

\def\zpc#1#2#3{    {\it Z. Phys. }{\bf C#1}:#2 (#3)}

\begin{document}

\title {{~}\\
{\Large \bf Recent experimental data and the size of
 }\\
{ ~}\\
{\Large \bf the  quark  in the Constituent Quark Model
}}

\author{ {~}\\
{~}\\
{\large\bf S. ~B o n d a r e n k o\,\,${}^{a)}$\,\thanks{Email: serg@post.tau.ac.il},
 \,  E.~L e v i n\,${}^{a)}$\,\thanks{E-mail: leving@post.tau.ac.il} 
\, and \,
 J.~N u i r i\,${}^{b)}$\,\thanks{E-mail: nyiri@rmki.kfki.hu} }
 \\[10mm]
 {\it\normalsize ${}^{a)}$ HEP Department,  School of Physics and Astronomy,}\\
 {\it\normalsize Raymond and Beverly Sackler Faculty of Exact Science,}\\
 {\it\normalsize Tel-Aviv University, Ramat Aviv, 69978, Israel}\\[0.5cm]
{\it\normalsize ${}^{b)}$ Research Institute for Particle and Nuclear
  Physics,}\\
 {\it\normalsize H-1525,  Budapest, P.O.B. 49,  Hungary }\\[0.5cm]
}
\maketitle
\thispagestyle{empty}

\begin{abstract}

We use the Constituent Quark Model (CQM) to describe
 CDF data on double parton cross section and HERA data
on the $\,J\,/\,\Psi$ ratio cross section of elastic
and inelastic diffractive productions. Our estimate shows that the
radius of the constituent quark turns out to be rather
small, $R^2_{quark}\,\approx\,0.1\,Gev^{-2}$, in accordance
with the assumption on which CQM is based.

\end{abstract}
\begin{flushright}
\vspace{-17cm}
TAUP - 2700-2002 \\
\today
\end{flushright}
\thispagestyle{empty}
\newpage

\section{Introduction}
One of the most challenging problems of QCD is to find correct degrees of
freedom for high energy ``soft'' interaction. In other words, the question
is what set of quantum numbers diagonalizes the interaction matrix at high
energies. On the one hand, the observation of the diffraction production
in all ``soft'' processes including the
photoproduction \cite{DIFFREV}  is a direct experimental indication that
hadrons are not correct degrees of freedom. On the other hand, at short
distances we know  that colour dipoles are the correct degrees of freedom \cite{MU94}
(see also Refs. \cite{DOF} ).
Frankly speaking, these two facts exhaust our solid theoretical knowledge
on the subject.

In this paper we are going to examine an old model for the degrees of        
freedom at high energy: the Constituent Quark Model\cite{AQM}, in which      
the constituent quarks play the r\^ole of the correct degrees of freedom.    
In spite of the naivity of this model it works and describes a lot of
``soft'' data in the first approximation \cite{QUARKBOOK}.

Different theoretical arguments have also been expressed in favour of the  
existence of two sizes in hadrons, e.g. in instanton models of QCD vacuum  
\cite{SHURYAK} and it has been included as an essential ingredient in
the non-perturbative QCD approach for the high energy scattering
\cite{HEIDELBERG}. The CQM gives a constructive way to build an approach
which introduces two dimensional scales inside a hadron: the size of
the hadron, built of the constituent quarks and the size of the
constituent quark itself.

We wish to re-examine this model because of two beautiful pieces of the
experimental data.

\begin{figure}[h]
\begin{minipage}{12.0 cm}
\begin{center}
\epsfxsize=10cm
\epsfysize=5cm
\leavevmode
\hbox{ \epsffile{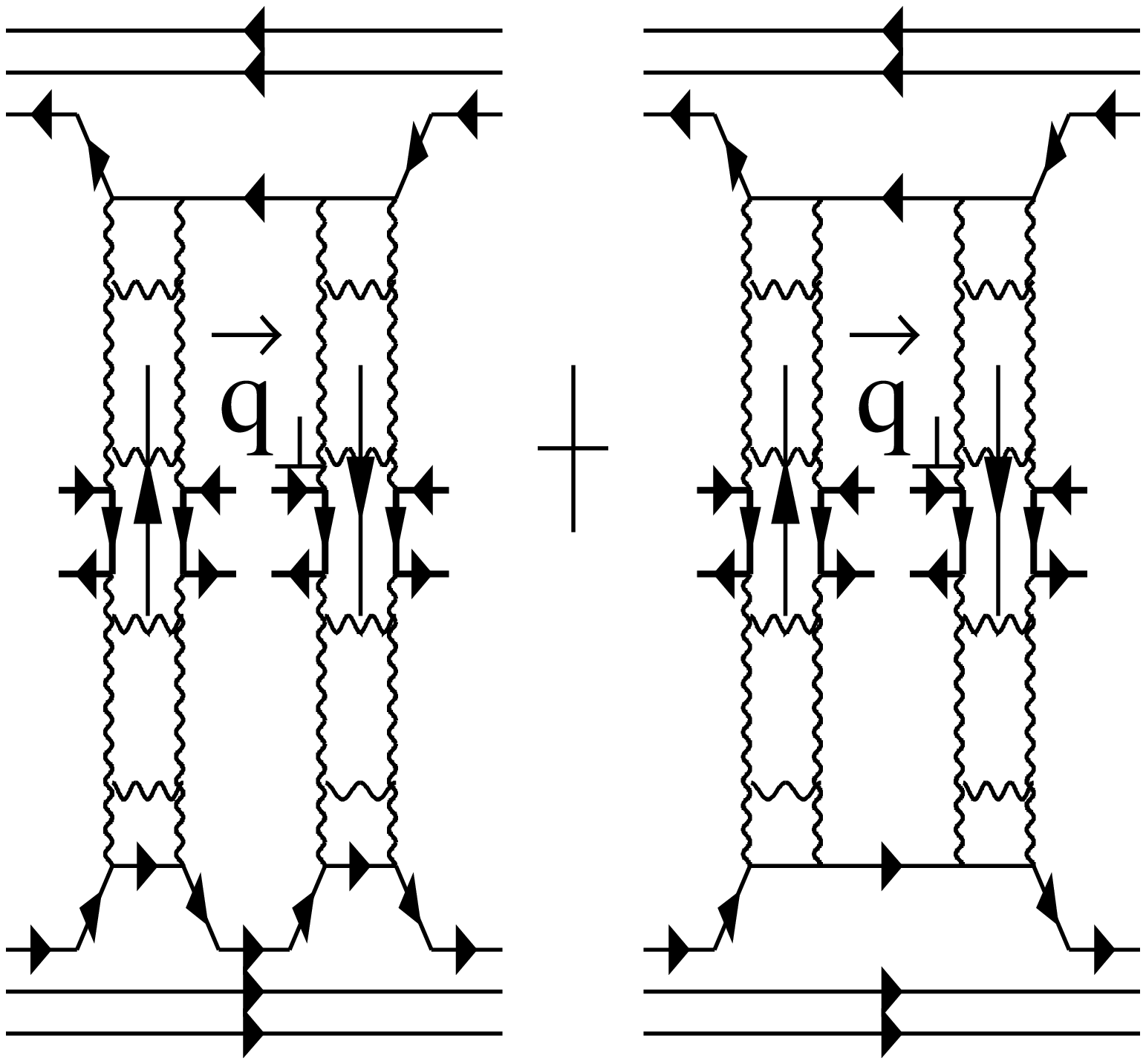}}
\end{center}
\end{minipage}
\begin{minipage}{4.0cm}
\caption{\it The Mueller diagram for the double parton shower interaction.}
\label{DP}
\end{minipage}
\end{figure}

{\it 1.\quad CDF double parton cross section at the Tevatron \cite{CDFDP}}.
The CDF collaboration has measured the process of inclusive production of
two pairs of ``hard'' jets with almost compensating transverse momenta in
each pair, and with values of rapidity that are very similar. Such pairs
can only be produced in double parton shower interactions (double parton
collisions, see Fig.~\ref{DP}). The cross section of this interaction can
be calculated using Mueller diagrams (as shown in Fig.~\ref{DP}) in CQM.

The double parton scattering cross section can be written in the form 
\cite{CDFDP}
\beq \label{DPXS}
\sigma_{DP}\,\,=\,\,m\,\frac{\sigma_{incl}( 2 \,jets)\,\,\sigma_{incl}( 2
  \,jets)}{2 \,\,\sigma_{eff}}\,\,,
\eeq
where the factor $m$ is equal to two for different pairs of jets and to
one for identical pairs. The experimental value of $ \sigma_{eff}
\,\,=\,\,14.5\,\pm\,1.7 \,\pm\,2.3\,\,\,mb$ \cite{CDFDP}. This value is
about 6 $\div$ 7 times less than the total $p \bar p $ cross section at
the Tevatron and this itself shows that we have a small 
scale inside the proton. The idea is that this small size is related to
the proper size of the constituent quark. We are going to check this idea
in the paper.

{\it 2. \quad HERA data on inclusive diffraction production with nucleon
excitation}.

\begin{figure}[h]
\begin{minipage}{12.0 cm}
\begin{center}
\epsfxsize=10cm
\leavevmode
\hbox{ \epsffile{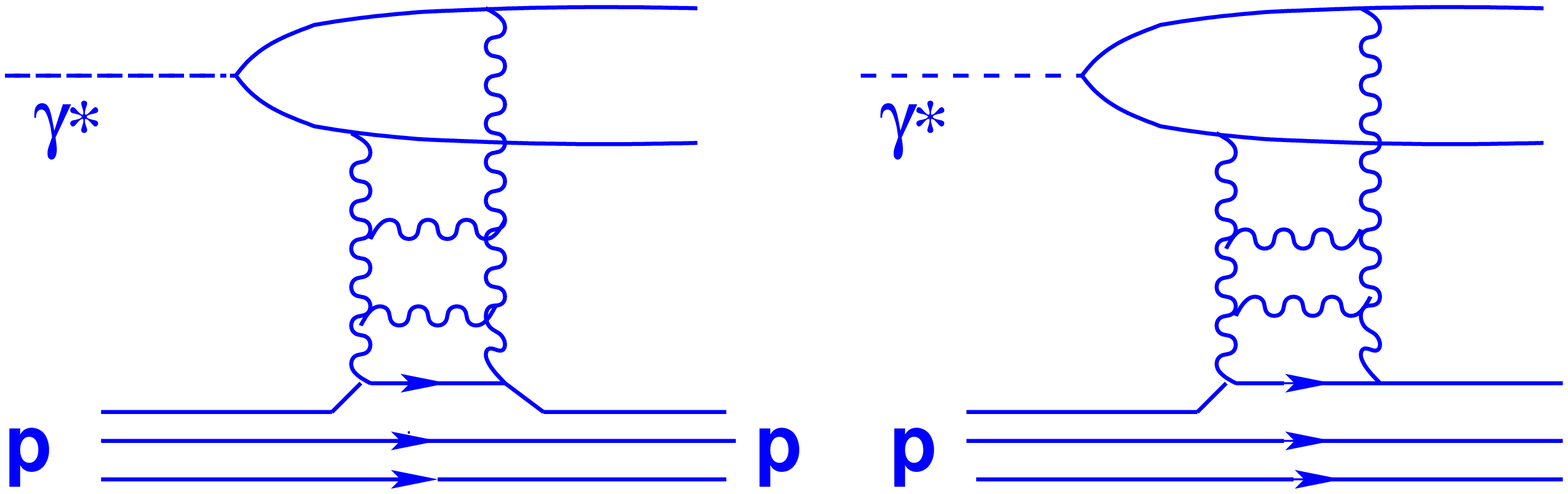}}
\end{center}
\end{minipage}
\begin{minipage}{4.0cm}
\caption{\it The inclusive diffraction in DIS in CQM.}
\label{DDIS}
\end{minipage}
\end{figure}

HERA data for both inclusive 
and exclusive diffractions, show that the nucleon excitations give at
least 30\% - 40\% of the cross section \cite{HERAREV} in the region of
small $t$ ($t < 1.5 \,GeV^2$). In CQM these two processes of the
diffraction production are presented in  Fig.~\ref{DDIS}. One can see that
they also give an information about the size of the constituent quark.

The main goal of this paper is to extract the value of the proper size of
the quark using these two sets of data, using the simplest assumptions on
the quark-quark interaction. We assume that (i) only the Pomeron exchange
\cite{POM} contributes to the amplitude of the quark-quark interaction at
high energies, and (ii) that we can calculate the inclusive cross
section using the Mueller diagrams \cite{MUDI} and the AGK cutting rules
\cite{AGK}.

In Section 2 we discuss in more detail our approach; we calculate the
value $\sigma_{eff}$ in \eq{DPXS} and the ratio
\beq \label{RA}
R(t)\,\, =\,\,\frac{\frac{d \sigma^{DIS}_{el}(\gamma^* + p \rightarrow X +  p)}{ d t}}{\frac{d \sigma^{DIS}_{inel}(  \gamma^* + p \rightarrow X
 + N^*)}{d t}}
\eeq
for the single Pomeron exchange model. In Section 3 we introduce the
possibility of triple Pomeron interactions and re-analyze $\sigma_{eff}$
and $R$. We present our conclusions and suggestions for further
experiments in the last section.
\section{Single Pomeron exchange in CQM.}

\subsection{Quark - quark scattering in the Pomeron approach}

As we said, the key ingredient of the CQM is the quark-quark
(antiquark) amplitude at high energies. In the single Pomeron model this
amplitude can be written in terms of the single Pomeron exchange as shown
in Fig.~\ref{qqpom}:

\begin{figure}[h]
\begin{minipage}{12.0 cm}
\begin{center}
\epsfxsize=10cm
\epsfysize=5cm
\leavevmode
\hbox{ \epsffile{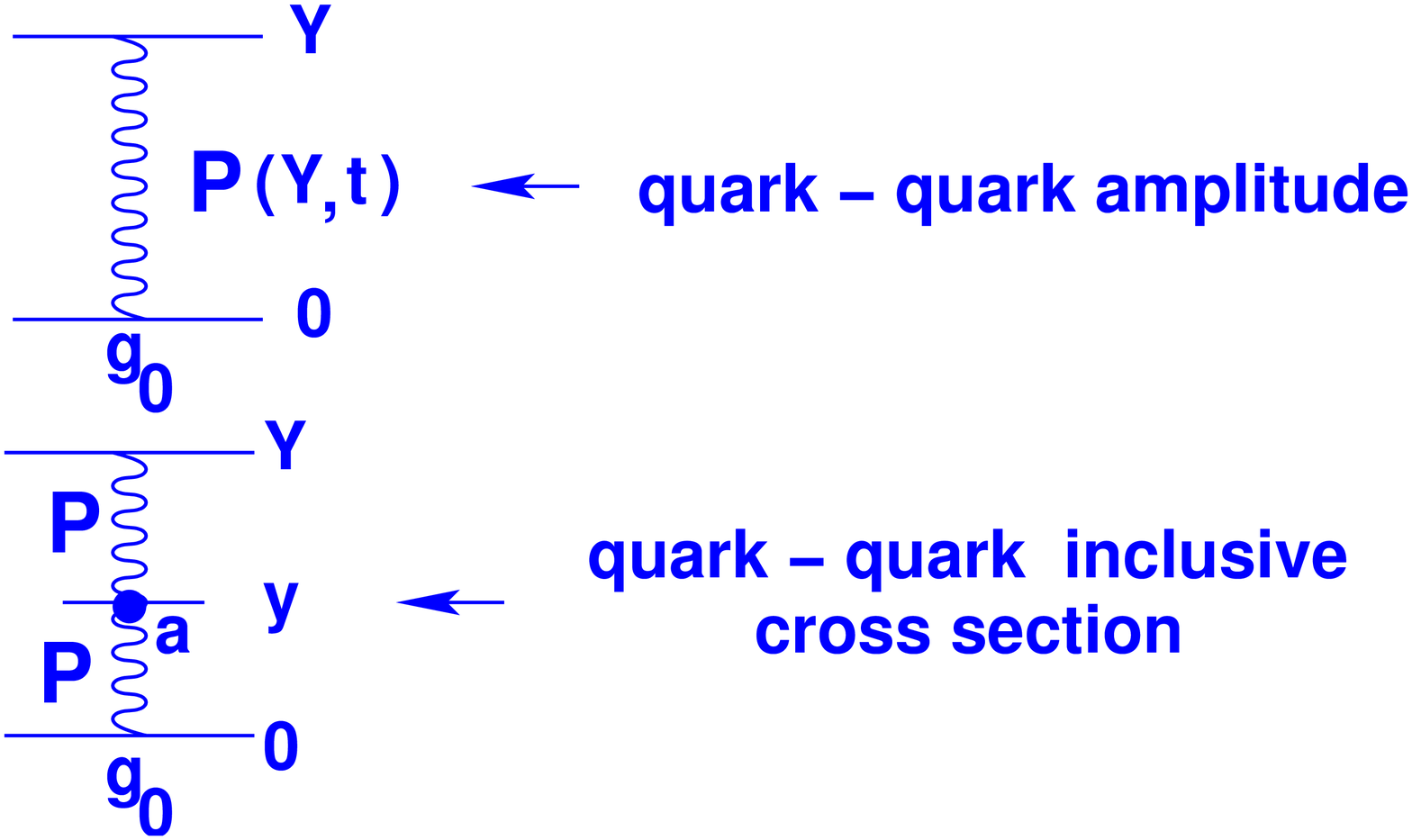}}
\end{center}
\end{minipage}
\begin{minipage}{4.0cm}
\caption{\it The quark - quark amplitude and the quark - quark inclusive
  cross section in the CQM.}
\label{qqpom}
\end{minipage}
\end{figure}
In order to calculate the contribution of the diagrams in Fig.~\ref{qqpom},
we have to know the main parameters of the Pomeron which we choose in the
following way.

\begin{itemize}
\item \quad  For the exchange of the ``soft'' Pomeron we have
\beq \label{Pom}
 P\Le Y, q^2\Ra=e^{\Delta Y-\alpha '\,\,q^2\,Y},
\eeq
where $Y=\ln s$ is the rapidity of the elastic process.

\item \quad We found the intercept and the slope of the Pomeron trajectory
by fitting the data of the total, elastic and diffractive cross sections
(we will show this later).
\beq \label{PARPOM}
\Delta=0.08\,-\,0.09\,;\quad\,\quad
\alpha'_P(0) \,\,\approx\,\,0.2\,\,GeV^{-2}\,\,;\quad
\quad \alpha_P(0) \,=\,1\,\,+\,\,\Delta
\eeq

\item \quad The vertices of the Pomeron-Quark interactions
\beq
g^{2}_{P-Q}=g^{2}_{0}\,\approx\,8\,-\,9\,\, GeV^{-2}.
\eeq
\item \quad The vertex for the inclusive emission of the hadron ($a$ in
  Fig.~\ref{qqpom}) is taken from the inclusive proton-proton scattering
  to be equal
\beq \label{AINCL}
a\,\,\approx \,\,2.
\eeq

\item \quad  It is common in the two processes we are going to discuss
  here that they are both due to the exchange of the so-called ``hard''
  Pomeron (gluon ``ladder'' in perturbative QCD, as shown in Fig.~\ref{DP}
  and Fig.~\ref{DDIS}). We accept here a rather simplified way of
  describing such  a ``hard'' Pomeron. We assume the same formulae
  as for the ``soft'' one (see \eq{Pom} and \eq{PARPOM}, but
  $\Delta_H \,\,>\,\,\Delta_S$ given by \eq{PARPOM} and $ \alpha'_P(0)
  \,=\,0$.

\end{itemize}

We use a very simple model for the wave function of the constituent quark
inside a hadron, namely,
\beq \label{QWF}
\Psi=\frac{\alpha}{\pi\sqrt{3}}\,
e^{-\frac{\alpha}{2}\Le\sum\,x^{2}_{i}\Ra}
\eeq
where the constant $\alpha$ is connected to the electromagnetic radius of
proton $R^2_{electr}$ :
\beq
\alpha=1/R^2_{electr},
\eeq
and we take $R^2_{electr}=15.6\,\,GeV^{-2}$.

\subsection{$\mathbf{\sigma_{eff}}$ in the CQM}

\begin{figure}[hptb]
\begin{tabular}{ c c}
\psfig{file=  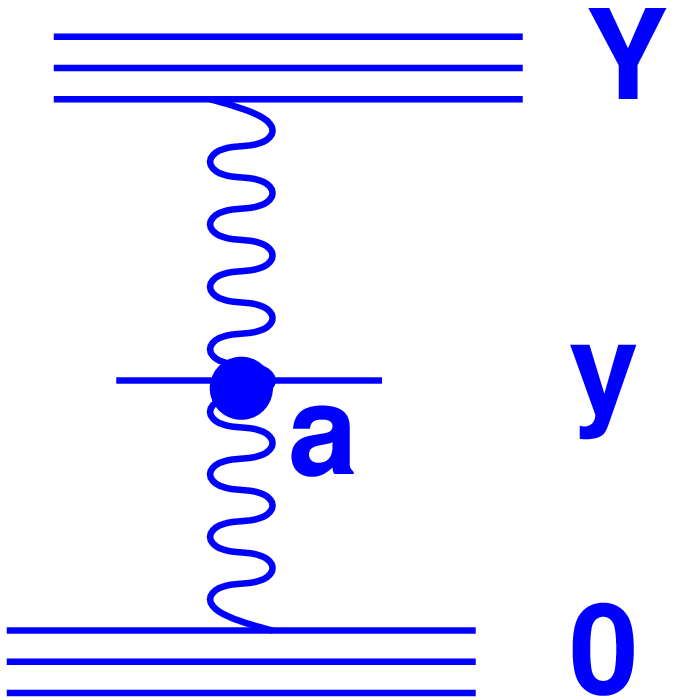,width=40mm,height=30mm} & \psfig{file=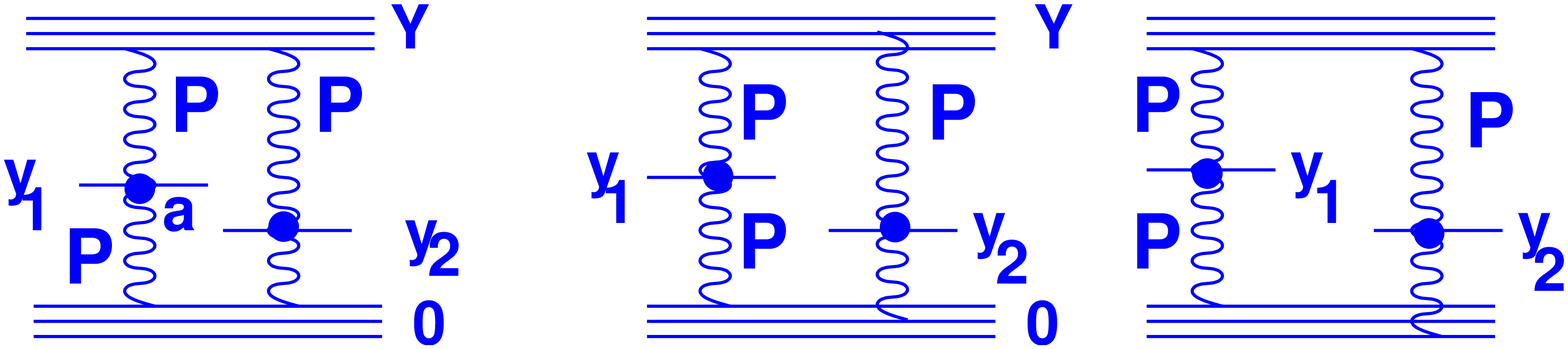,width=120mm,height=30mm}\\
       &  \\
Fig.4-a & Fig. 4-b \\
 &  \\
\end{tabular}
\caption{\it  The single inclusive (Fig. 4-a ) and double inclusive (Fig.4-b)
  cross sections in the single Pomeron exchange model in the CQM.}
\label{ppP}
\end{figure}

Armed with the knowledge that was discussed above, we can calculate the
contributions of the single Pomeron exchange to our processes
(see Fig.~\ref{ppP}).

In the case of the single inclusive process we have only one diagram,
of Fig.~\ref{ppP}-a,
all other contributions are cancelled
due to the AGK cutting rules \cite{AGK}. This diagram gives us
\beq \label{SINXSPP}
f_1(Y) \,\,=\,\,\frac{d \sigma}{d y} \,\,=\,\, 18\, g_{0}^2 \,a\,
\,e^{\Delta Y};
\eeq
we have here a combinatorial factor 9 for three quarks
in each proton, and an additional factor 2.
Indeed, we use the s-channel unitarity equation in the form:
\beq
2\,Im\,A_{el}\,=\,|\,A_{el}\,|^2\,+\,G_{in}\,\,  ,
\eeq
and we see that for the one Pomeron exchange, which is related to the
$G_{in}$ and therefore to the cut Pomeron, we have an additional 2.
This will be our usual prescription for the cut Pomeron.
For the double inclusive cross section we have contributions of several
diagrams, given in Fig.~\ref{ppP}-b.

\begin{enumerate}

\item Let us demonstrate, step by step, how to perform the calculations
in this model. We have for Fig.~\ref{ppP}-b :
\beq
D_{1}(y_1,y_2)\,=4\,\cdot\,9\,\int\,\frac{dt}{16\,\pi}\,
e^{2\,\Delta\,Y}\,e^{-|t|(4\,R^2_{quark}+2\alpha'_P\,Y)}\,.
\eeq
Here 9 is a combinatorial factor, 4 comes from the AKG rules (2 for each
cut Pomeron). We introduce also the constituent quark radius
$R^{2}_{quark}$ in order to take into account the $t$ - dependence of the
quark - Pomeron vertex, which we parametrize in the simplest Gaussian
form: $g_{q-P}(t) = g_0 \exp[R^2_{quark}\,t]$.
Integrating this expression over $|t|$, we obtain

\beq
D_{1}(y_1,y_2)\,=\,\frac{9}{8\,\pi}\,
\frac{g_0^4\,a^2\,\alpha \,e^{2\,\Delta\, Y}}{\,\alpha'_P\,Y\,+\,R^2_{quark}}.
\eeq

\item The second diagram of Fig.~\ref{ppP}-b is somewhat more complicated.
First of all, we define the form factor for this type of diagrams:
\beq
\int\,|\,\Psi\,\Le\,x_1\,x_2\,x_3\,\Ra\,|^2\,\delta\Le\,\vec{x}_1+
\vec{x}_2+\vec{x}_3 \,\Ra\,
e^{i\,q\,x_1\,-i\,q\,x_2}\,
dx_1\,dx_2\,dx_3\,=\,e^{-\frac{|t|}{2\,\alpha}}\,\,
\eeq
This gives
\beq
D_{2}(y_1,y_2)\,=
4\,\cdot\,36\,\int\,\frac{dt}{16\,\pi}\,
e^{2\,\Delta\,Y}\,e^{-|t|\,2\,(\,\alpha'_P\,Y\,+\,\frac{1}{2\,\alpha})}\,,
\eeq
where 36 is the combinatorial factor for this type of configurations,
and we obtain for this diagram:
\beq
D_{2}(y_1,y_2)\,=
\,\frac{9}{\pi}\,
\frac{g_0^4\,a^2\,\alpha \,e^{2\,\Delta\, Y}}{1+2\alpha\,\alpha'_P\, Y}.
\eeq
\item  The third diagram in Fig.~\ref{ppP}-b gives
\beq
D_{3}(y_1,y_2)\,=
4\,\cdot\,36\,\int\,\frac{dt}{16\,\pi}\,
e^{2\,\Delta\,Y}\,e^{-|t|\,(2\,\alpha'_P\,Y\,+
\,2\,R^{2}_{quark}+
\,\frac{1}{2\,\alpha})}\,.
\eeq
Here we used the same vertex, and after the integration we get
\beq
D_{3}(y_1,y_2) \,=\,\frac{9}{\pi}\,
\frac{g_0^4 \,a^2\,\alpha\,e^{2\,\Delta \,Y}}{1+4\alpha\,\Le\alpha'_P\, Y+
\,R^{2}_{quark}\Ra}.
\eeq
\end{enumerate}

Now we are ready to use our equation (~\ref{DPXS}) and to estimate
the possible dependence of $\sigma_{eff}$ on the quark radius
$R^{2}_{quark}$. For the $\sigma_{eff}$ we can write
\beq \label{SIEFF}
\sigma_{eff}\,\,=\,\,m\,\frac{ f(y_1)\,\,f(y_2)}{2\,\Le
D_{1}\Le R^{2}_{quark}\Ra\,+
\,D_{2}\Le R^{2}_{quark}\Ra\,+\,D_{3}\,\Ra}\,\,.
\eeq

\begin{figure}[hptb]
\begin{tabular}{ c c}
\psfig{file=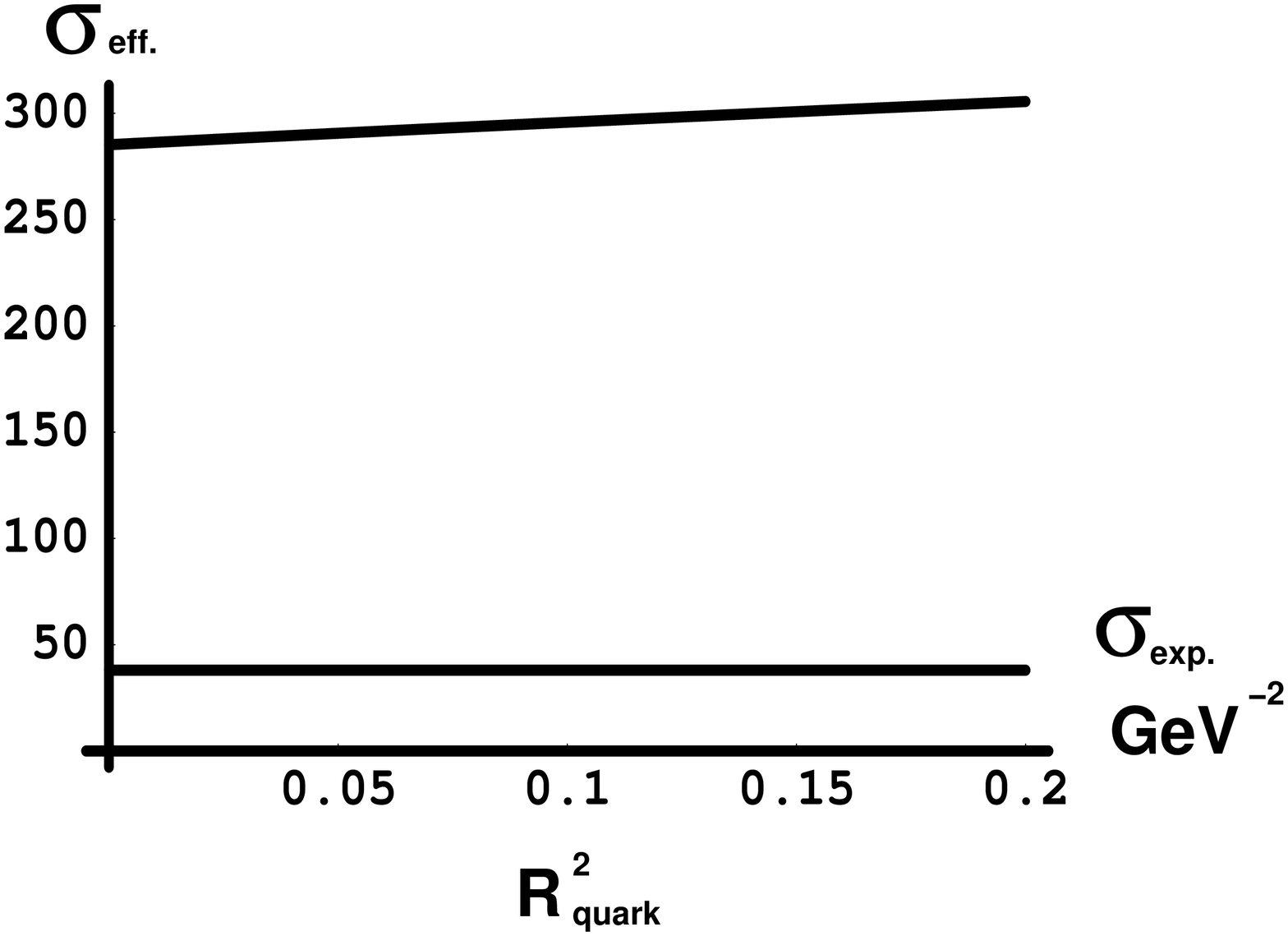,width=80mm} &
\psfig{file=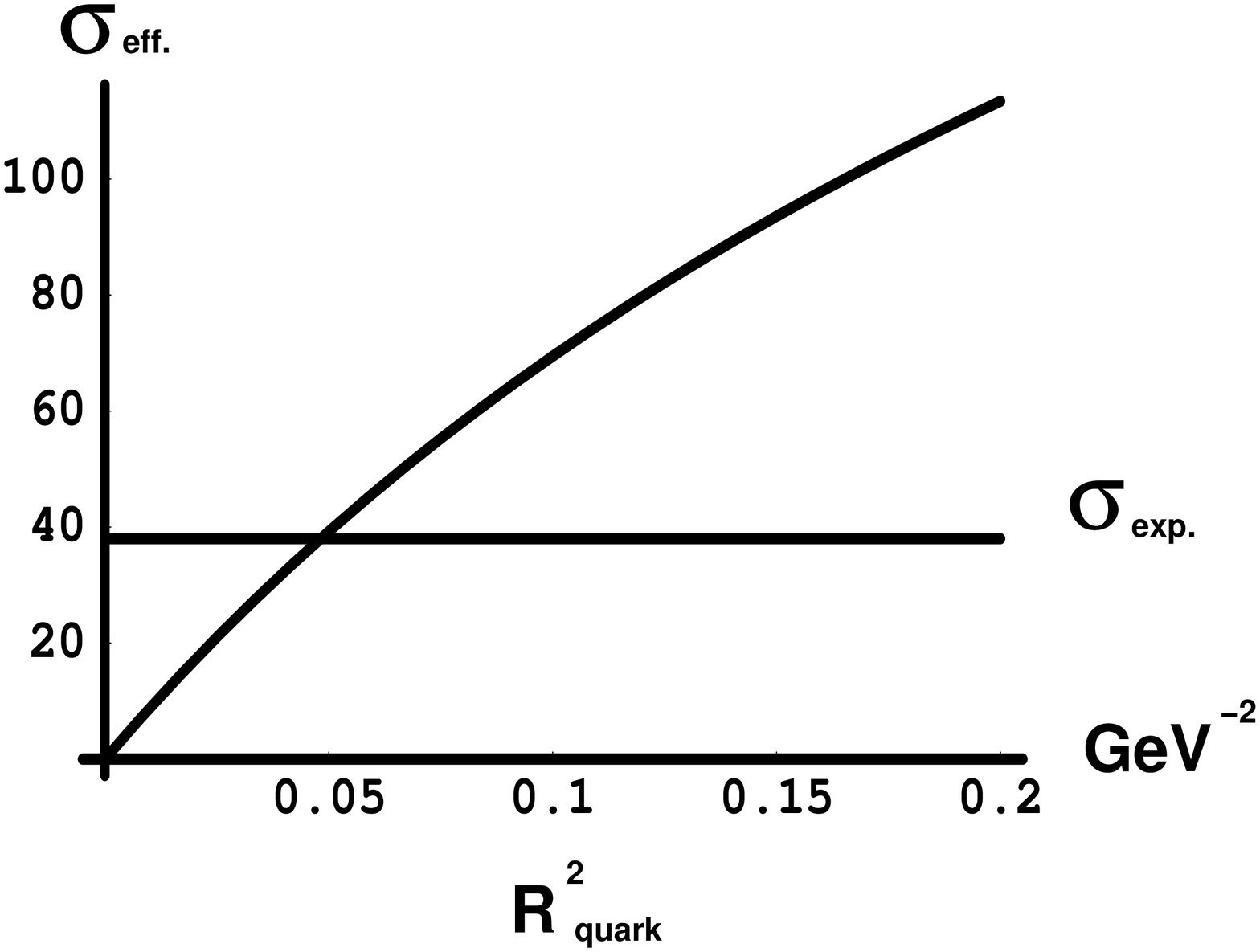,width=80mm}\\
       &  \\
Fig.5-a & Fig. 5-b \\
 &  \\
\end{tabular}
\caption{\it  $\sigma_{eff}$ (in $GeV^{-2}$)  versus $R^2_{quark}$ (in
  $GeV^{-2}$)  for $\alpha'_P =
  0.25\,GeV^{-2}$ (Fig.5-a) and for  $\alpha'_P = 0 $(Fig.5-b) .The straight
  line in Fig. 5-b shows the experimental value of $\sigma_{eff}$. }
\label{rat1}
\end{figure}

In Fig.~\ref{rat1} we plot the value of $\sigma_{eff}$ from \eq{DPXS} as
a function of $R^2_{quark}$ for $m=1$, $y_1=y_2=Y/2$.
One can see that at any value of the unknown size of the constituent
quark the value of $\sigma_{eff}$ turns out to be larger than the
experimental value.

The conclusion we derive from this simple model is quite obvious: in CQM
the  ``soft'' Pomeron exchange for quark-quark scattering cannot
explain the CDF result of the double parton cross section. However,
the experimental value of $\sigma_{eff}$ obtained from the high $p_t$
jet production can be described by a ``hard'' Pomeron. Indeed, in this
case we have to consider $\alpha'_P = 0$ and Fig.~\ref{rat1}-b shows that
we have $\sigma_{eff} \,\approx\,15\,mb$ for
$R^2_{quark} \,\approx 0.05\,GeV^{-2}$. We would like to emphasize that
the double parton shower cross section $ \sigma_{eff}$ is very sensitive
to the size of the constituent quark in the CDF kinematics since the CDF
measures the jet production by ``hard'' Pomeron for which
$\alpha'_P(0) = 0$.

\subsection{$\mathbf{R(t)\,\, =
\,\,\frac{d\sigma^{DIS}_{el}/dt}{d\sigma^{DIS}_{inel}/dt}\,\,\,in\,\,\,
the\,\,\, CQM}$}

\begin{figure}[h]
\begin{minipage}{12.0 cm}
\begin{center}
\epsfxsize=10cm
\leavevmode
\hbox{ \epsffile{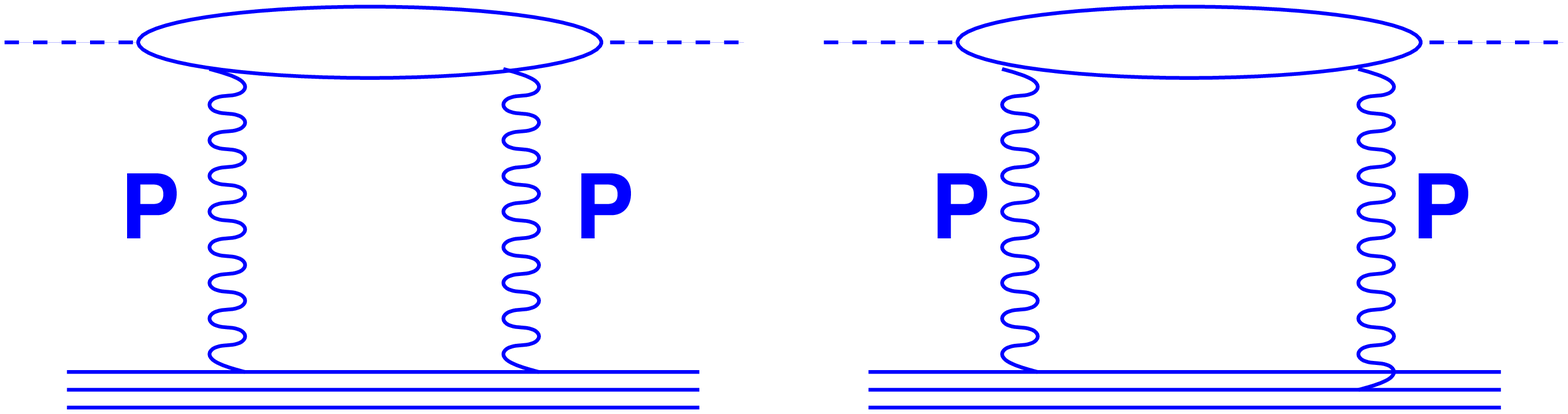}}
\end{center}
\end{minipage}
\begin{minipage}{4.0cm}
\caption{\it The total inclusive diffraction cross section for a single
  Pomeron exchange in the CQM.}
\label{gapom}
\end{minipage}
\end{figure}

In the single Pomeron exchange model the cross section for DIS
diffraction and elastic scattering can be described by two diagrams in
Fig.~\ref{gapom}.

The first diagram in Fig.~\ref{gapom} leads to
\beq \label{DISP1}
\frac{d\sigma^D_1}{d t}\,\,=\,\,3 g^2_0\,e^{2\,\Delta\,Y}\,e^{- 2|t|
( \alpha'_H\,Y + R^2_{quark})}\,\,.
\eeq

In this process the ``hard'' Pomeron contributes, so
$\Delta=\Delta_H\,\approx\,0.3$ and $\alpha'_H(0) = 0$.
It should also be stressed that the contribution of
this diagram to the total cross section (integrated over $t$) is very
sensitive to the value of $R_{quark}$ ($D_1\,\,\propto \,\,1/R^2_{quark}$)
since this contribution is proportional to $1/(\alpha'_P\,Y
\,\,+\,\,R^2_{quark})$, and for the ``hard '' Pomeron $\alpha'_P(0) = 0$.

\begin{figure}[hptb]
\begin{center}
 \psfig{file=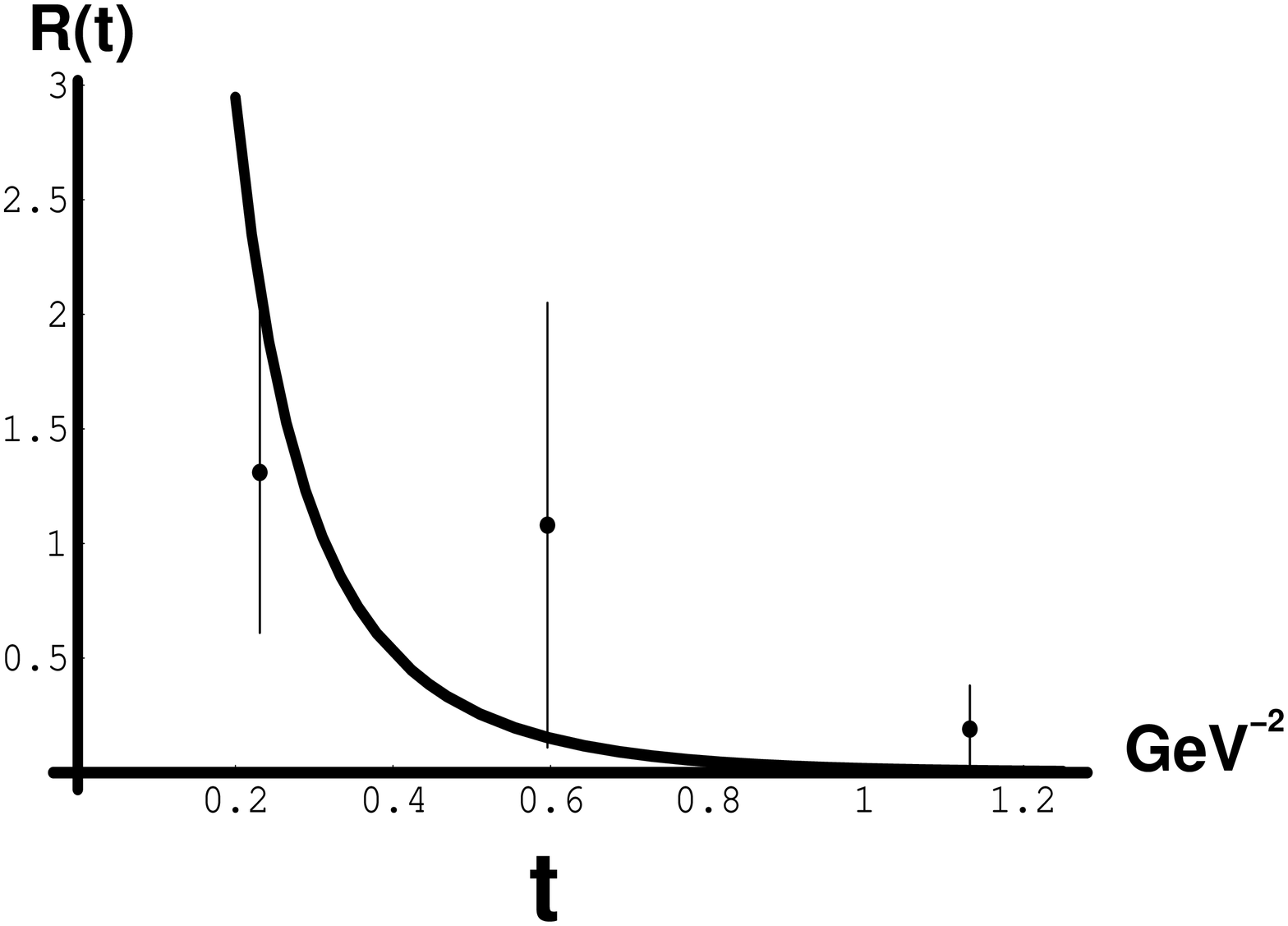,width=80mm}\\
 \end{center}
\caption{\it  $R(t)$ versus $t$, experimental data were taken
from \cite{DIFFREV}.}
\label{rat2}
\end{figure}

The second diagram in Fig.~\ref{gapom} 
depends only slightly on the value of the quark radius, and can be
described as follows:

\beq \label{DISP2}
\frac{d\sigma^D_2}{d t}\,\,=\,\,6 g^2_0\,e^{2\,\Delta\,Y}\,e^{-|t|
  \Le\,\frac{1}{2 \alpha}\, +\,2\,
  (\,\alpha'_H \,Y\,\,+\,\,R^2_{quark})\,\Ra}\,\,.
\eeq

For the elastic cross section we have
\beq \label{DISP3}
\frac{d\sigma^{el}_2}{d t}\,\,=\,\,9 g^2_0\,e^{2\,\Delta\,Y}\,e^{-|t|
  \Le\,\frac{1}{3 \alpha}\, +\,2
  (\,\alpha'_H \,Y\,\,+\,\,R^2_{quark})\,\Ra}\,\,.
\eeq
Here we used the expression for the simplest form factor
in this model:
\beq
\int\,|\,\Psi\,\Le\,x_1\,,x_2\,,x_3\,\Ra\,|^2\,\delta\Le\,
\vec{x_1}+\vec{x_2}+\vec{x_3}\,\Ra\,
\,e^{i\,q\,x_1}\,
dx_1\,dx_2\,dx_3\,=\,e^{-\frac{|t|}{6\,\alpha}}\,\,.
\eeq
Now we define our $d\sigma^{DIS}_{inel}/dt$ as a difference between
these two contributions. Fig.~\ref{rat2} shows the value of the ratio
$R(t)$ , given by \eq{RA}. One can easily see that this ratio
depends neither on the value of $R_{quark}$, nor on the value of the
``hard'' Pomeron intercept $\Delta$. Hence, it can be considered as a
crucial test of our model.

Comparing two pictures:  Fig.~\ref{rat1} and Fig.~\ref{rat2}, we conclude
that $\sigma_{eff}$ can be described using the small value of $R_{quark}$ and
the ``hard'' Pomeron approach with $\alpha'_P = 0 $, while $R$ does
not depend on $R_{quark}$ and $\alpha'_P = 0 $, and the obtained shape
of R for a given t is close to the experimental graph in \cite{DIFFREV}.
It follows from this simple exercise that we have to check the
description of high energy scattering in the CQM including the "triple"
Pomeron vertex. But before doing this, in order to clarify the question
of the possible value of triple Pomeron vertex and other parameters
in this model, we will examine our model by fitting the data on total,
elastic and diffractive dissociation cross sections.

\subsection{Total, elastic and diffractive dissociation processes in
the CQM}

Let us check, how well our model, CQM, describes experimental data on
total, elastic and diffractive dissociation processes. In the simplest
case, without the triple Pomeron vertex corrections and additional Pomeron
exchanges, we have the following contribution for the total cross section:

\beq
\sigma_{total}\Le\,Y\,\Ra=
18\,g_0^2\,e^{\Delta\,Y}\,\,.
\eeq
Due to our s-channel unitarity constraint we write here again an
additional factor two for the cut Pomeron. For the elastic cross
section one can obtain

\beq
\sigma_{Elastic}\Le\,Y\,\Ra=
\frac{81}{16\,\pi}\,\frac{g_0^4\,e^{2\,\Delta\,Y}}
{2\,\alpha'_P\,Y\,+\,4\,R^2_{quark}+2/(3\,\alpha)}\,\,.
\eeq

We define also how to calculate the diffractive dissociation process.
First of all, we take into account the sum of diffractive and elastic
processes. We have to calculate several types of diagrams, see
Fig.~\ref{Elad}. The answer for these diagrams is the following:

\begin{figure}[hptb]
\begin{center}
 \psfig{file=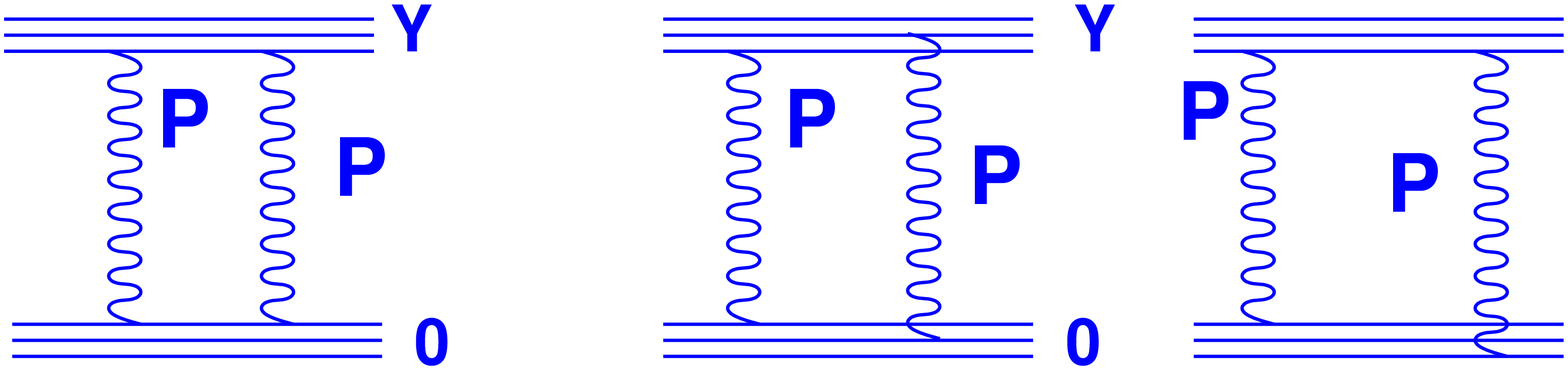,width=100mm}\\
 \end{center}
\caption{\it Diagrams for diffractive and elastic cross sections
in the first order .}
\label{Elad}
\end{figure}

\beq
\sigma\,\Le\,Y\,\Ra=
\frac{9}{32\,\pi}\,\frac{g_0^4\,e^{2\,\Delta\,Y}}
{\alpha'_P\,Y\,+\,2\,R^2_{quark}}\,+\,
\frac{9}{4\,\pi}\,\frac{g_0^4\,\alpha\,e^{2\,\Delta\,Y}}
{1+2\,\alpha\,\alpha'_P\,Y\,}\,+\,
\frac{9}{2\,\pi}\,\frac{g_0^4\,\alpha\,e^{2\,\Delta\,Y}}
{1+4\,\alpha\,(\alpha'_P\,Y\,+\,R^2_{quark})}\,.
\eeq

The experimental value of the diffractive dissociation cross section,
single and double together, is equal to:

\beq
\sigma_{Diffr}\,\Le\,Y\,\Ra\,=\,
\sigma\Le\,Y\,\Ra\,-
\sigma_{Elastic}\Le\,Y\,\Ra\,\,.
\eeq

\begin{figure}[hptb]
\begin{center}
 \psfig{file=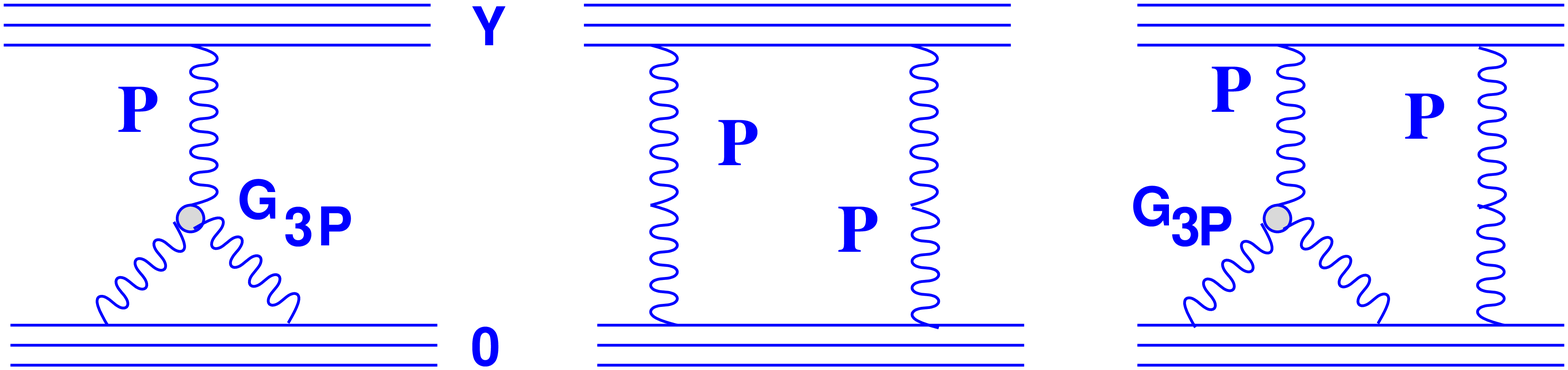,width=110mm}\\
 \end{center}
\caption{\it Higher order corrections.}
\label{Next}
\end{figure}

\begin{figure}[hptb]
\begin{tabular}{ c c}
\psfig{file=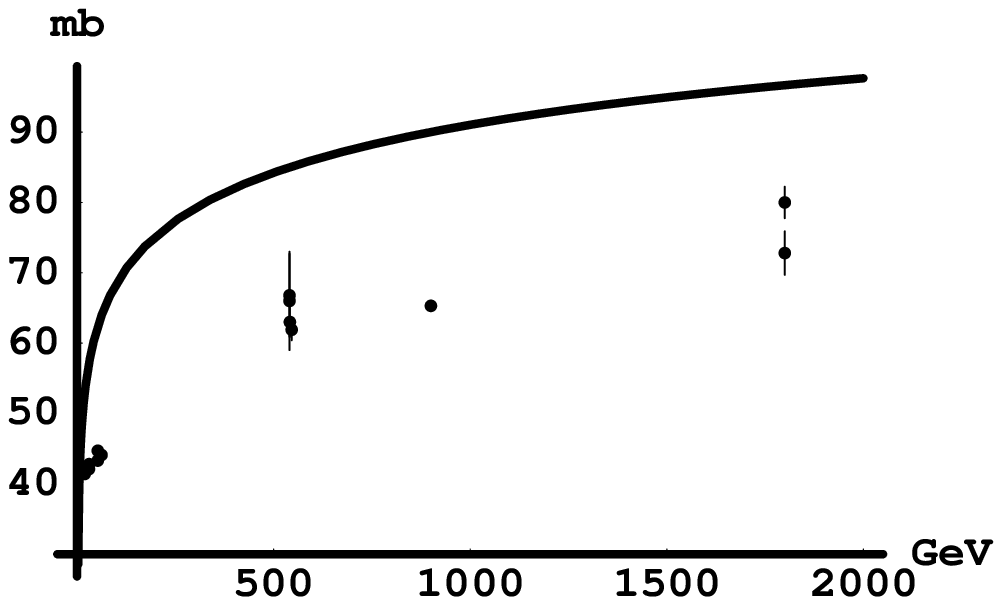,width=80mm} &
\psfig{file=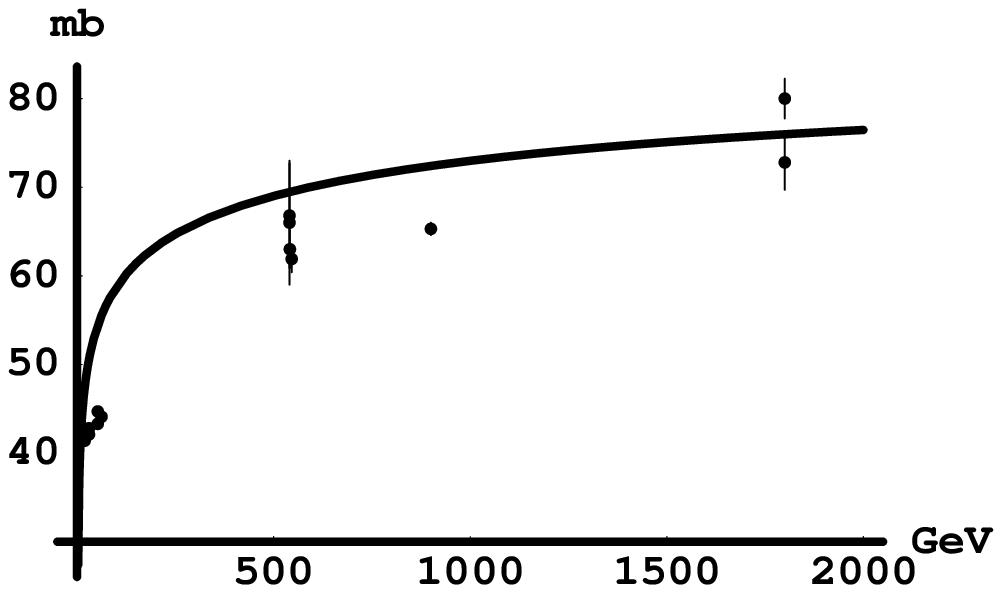,width=80mm}\\
       &  \\
Fig.10-a & Fig. 10-b \\
 &  \\
\end{tabular}
\caption{\it
Total cross section of p-p interaction in CQM
without $\gamma$, Fig.10-a, and with
$\gamma$ corrections, Fig.10-b. }
\label{Tot0}
\end{figure}

\begin{figure}[hptb]
\begin{tabular}{ c c}
\psfig{file=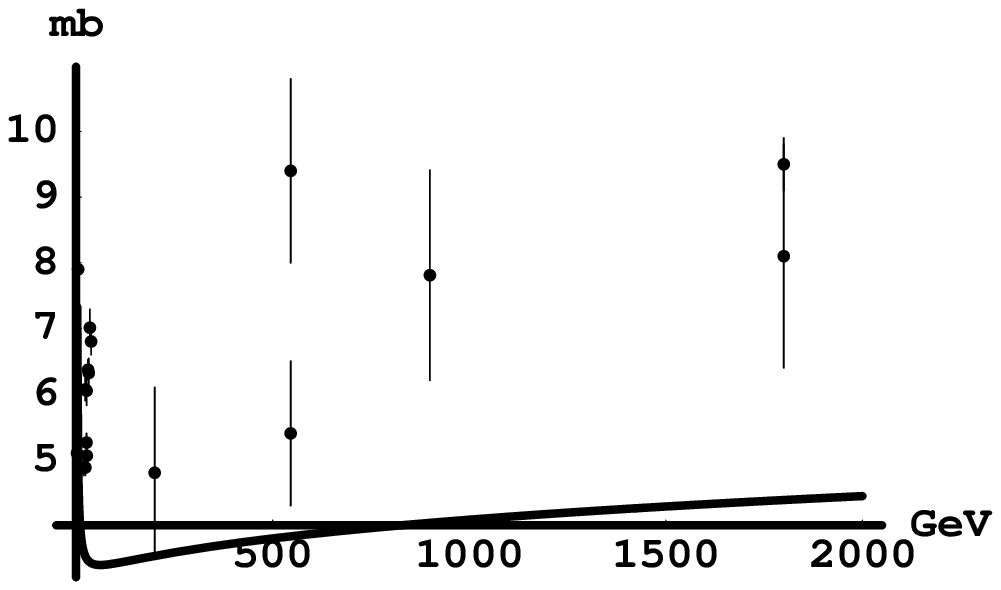,width=80mm} &
\psfig{file=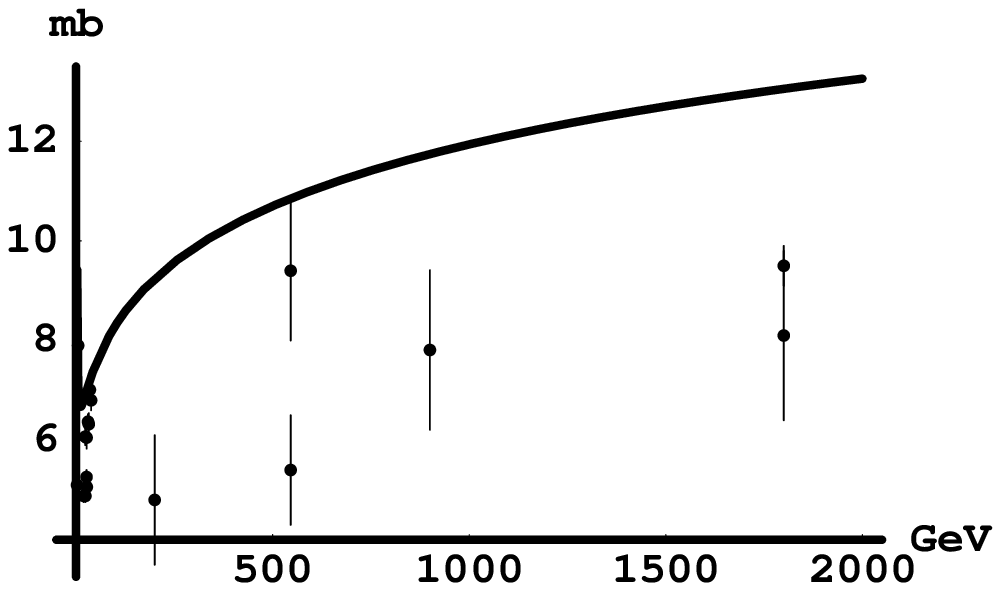,width=80mm}\\
       &  \\
Fig.12-a & Fig. 12-b \\
 &  \\
\end{tabular}
\caption{\it
Diffractive dissociation cross section of p-p interaction in CQM
without $\gamma$, Fig.12-a, and with
$\gamma$ corrections, Fig.12-b. }
\label{Diffr}
\end{figure}

The result of the fitting is presented in Figs.~\ref{Tot0}\,-\,
~\ref{Diffr}. We see that we have to include into our consideration the
diagrams of the next order, with the triple Pomeron vertex and with the
double Pomeron exchange. Taking into account these corrections, which are
given by diagrams of the type of Fig.~\ref{Next}, we obtain better a
fit, see Figs.~\ref{Tot0}\,-\,~\ref{Diffr}.

We present the analytical expressions for these contributions in the
Appendix; they are rather simple. Let us note also that we calculated
both the single and double diffractive dissociation processes, and we
plotted a graph with single diffractive dissociation data. Hence, it is
not surprising that the obtained curve is above the experimental points.
So, in this model we obtained the following parameters: for the
Pomeron-quark vertex and intercept
$g_{P-Q}^2\,=\,8\,-\,9\,GeV^{-2}$ ,$\Delta\,=\,0.08\,-\,0.09\,$,
and for the value of the triple Pomeron vertex
$\frac{\gamma}{\g_0}\,=\,0.013\,-\,0.015\,$.
The conclusion is 
that the value of the triple Pomeron vertex is not small and the
contribution of next order corrections is important.

\begin{figure}[hptb]
\begin{tabular}{ c c}
\psfig{file=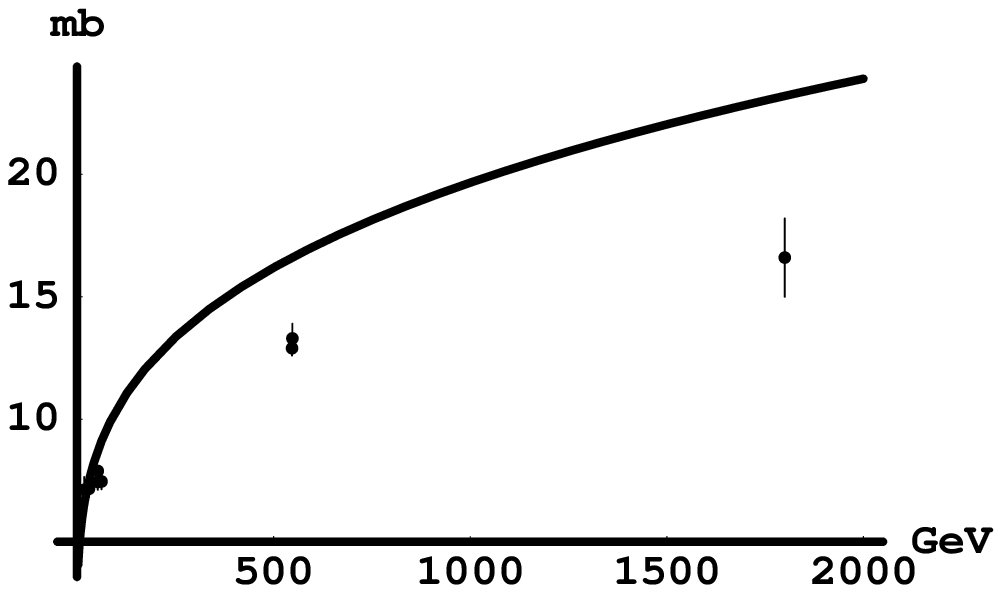,width=80mm} &
\psfig{file=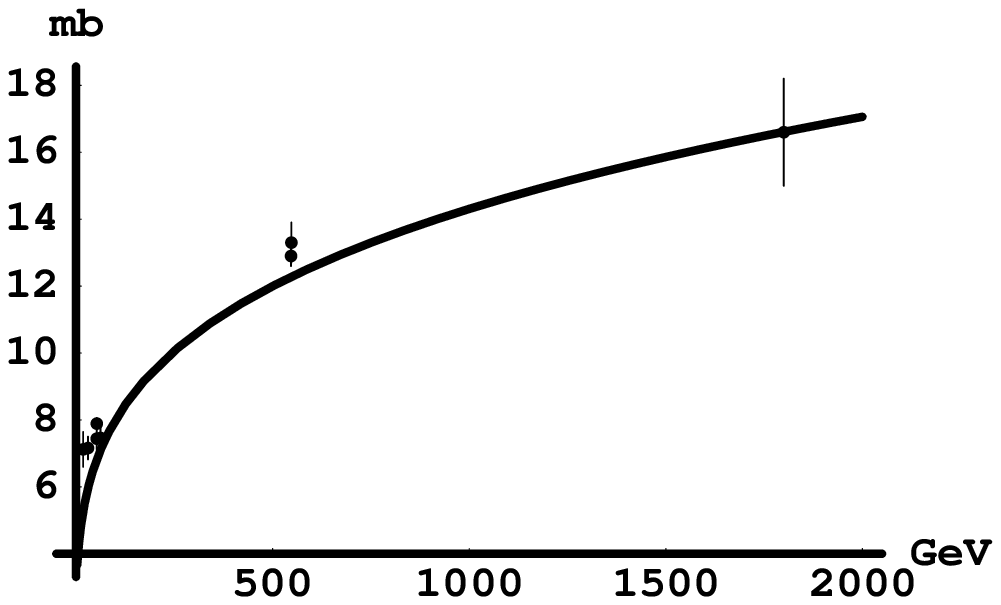,width=80mm}\\
       &  \\
Fig.11-a & Fig. 11-b \\
 &  \\
\end{tabular}
\caption{\it
Elastic cross section of p-p interaction in CQM
without $\gamma$, Fig.11-a, and with
$\gamma$ corrections, Fig.11-b. }
\label{Ela0}
\end{figure}

\section{Triple Pomeron interaction in CQM}

In this section we are going to calculate the same processes as in the
previous section, but taking into account the triple Pomeron interaction. We
would like to see how this interaction changes the $R_{quark}$ dependence
of the amplitude, and we will try to find a reasonable value of the quark
radius to be used in the CQM based phenomenology.

\subsection{$\mathbf{\sigma_{eff}}$ in one loop calculation}
\begin{figure}[hptb]
\begin{center}
 \psfig{file=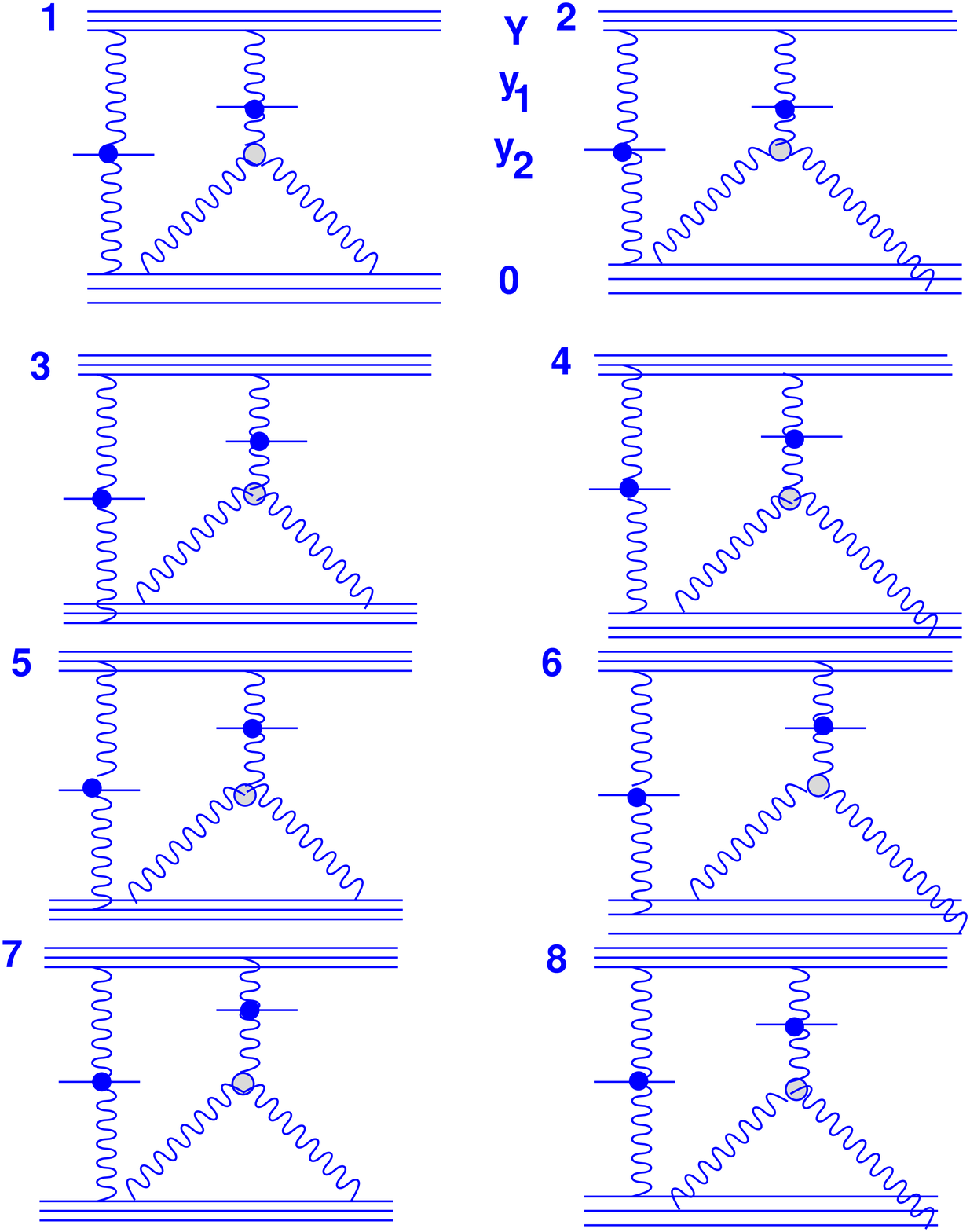,width=160mm}\\
 \end{center}
\caption{\it  Diagrams incorporating the triple Pomeron interaction
  for $\sigma_{eff}$.}
\label{pp3P}
\end{figure}

The triple Pomeron interaction leads to a number of new diagrams presented in
Fig.~\ref{pp3P} which we have to add to the Fig.~\ref{ppP} diagrams.
It should be stressed that there are many diagrams which, due to the AGK
cancellations \cite{AGK}, do not contribute to double inclusive
production.

\begin{figure}[hptb]
\begin{center}
 \psfig{file=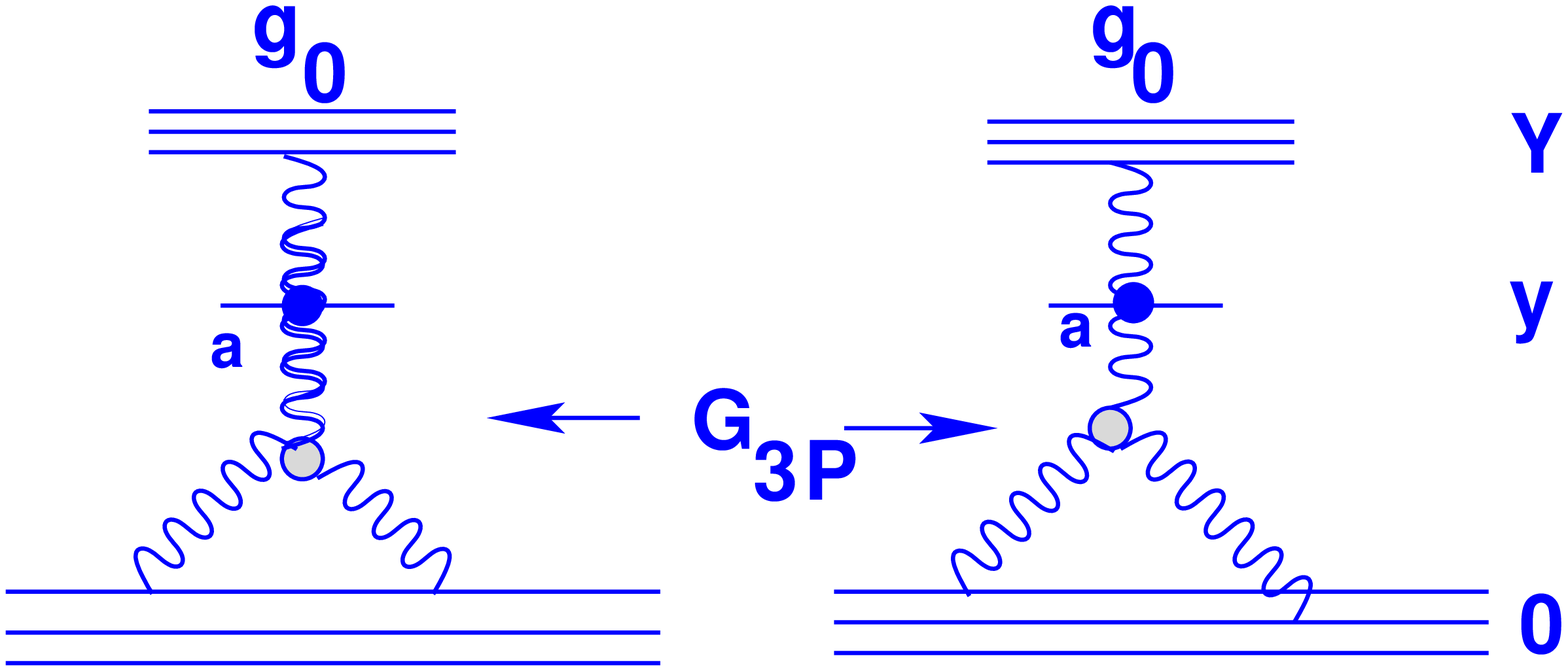,width=110mm}\\
 \end{center}
\caption{\it  The diagrams with $\gamma$
contributing to the single inclusive process.}
\label{singl1}
\end{figure}

The diagrams with triple Pomeron interaction in the case of a single
inclusive process are presented in Fig.~\ref{singl1}. These diagrams are
the simplest ones with the triple Pomeron vertex, which we denote by
$\gamma$, and where we take $y_1=y_2=Y/2$. So, we have the following
additional contributions to the single inclusive process.

\begin{enumerate}

\item The first diagram of Fig.~\ref{singl1} is

\beq
f_{2}(Y)\,=\,\frac{9}{8\,\pi}\,
\,g_0^4\,e^{\Delta\, Y}\,\Le\frac{\gamma}{g_0}\Ra\,
\int_{0}^{Y/2}\,\frac{dx\,e^{\Delta\,x}}{R_{quark}^2+\alpha'_P\, x}.
\eeq

\item The second diagram of Fig.~\ref{singl1} gives

\beq
f_{3}(Y)\,=\,\frac{9}{\pi}\,
\,g_0^4\,e^{\Delta\, Y}\,\alpha\,\Le\frac{\gamma}{g_0}\Ra\,
\int_{0}^{Y/2}\,\frac{dx\,e^{\Delta\,x}}{1+4\,\alpha\,\alpha'_P\, x}.
\eeq

\end{enumerate}

We have more complicated contributions for the double inclusive cross
section. Indeed, these are all diagrams of Fig.~\ref{pp3P} and
Fig.~\ref{doub1}. The analytical expressions for these diagrams are
given in the Appendix.

\begin{figure}[hptb]
\begin{center}
 \psfig{file=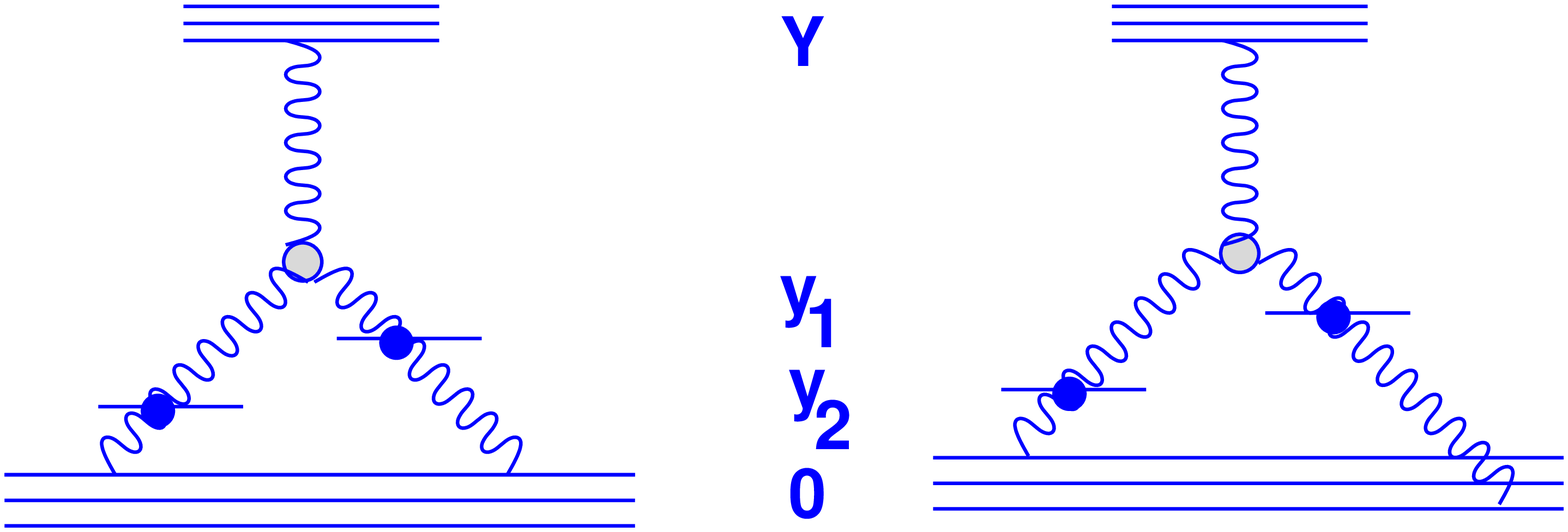,width=120mm}\\
 \end{center}
\caption{\it  The simplest diagrams with $\gamma$
contributing to the double inclusive process.}
\label{doub1}
\end{figure}

Now, using \eq{DPXS} again, we estimate the dependence
of $\sigma_{eff}$ on $R_{quark}^2$ in the case of one Pomeron loop.
We have:

\beq \label{SIEFF1}
\sigma_{eff}\,\,=\,\,m\,
\frac{\Le f_1(y_1)\,+\,f_2(y_1)\,+\,f_3(y_1)\Ra\Le
f_1(y_2)\,+\,f_2(y_2)\,+\,f_3(y_2)\Ra}
{2\,\Le
D_{1}+D_{2}+D_{3}+D_4+D_5+
D_{6}+D_{7}+D_{8}+D_{9}+
D_{10}+D_{11}+D_{12}+D_{13}\Ra}\,.
\eeq

\begin{figure}[hptb]
\begin{center}
\psfig{file=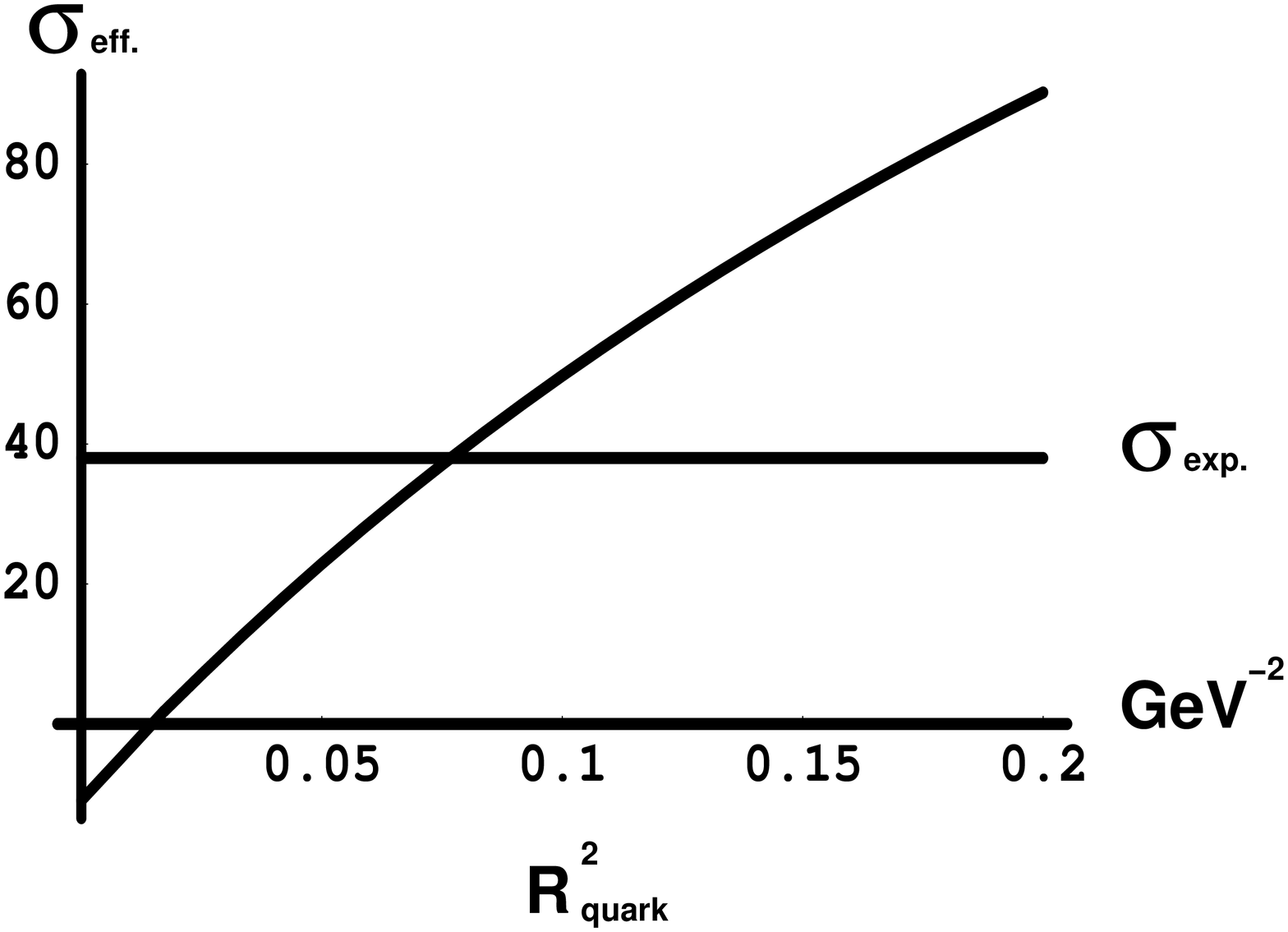,width=80mm}
\end{center}
\caption{\it  $\sigma_{eff}$ (in $GeV^{-2}$)  versus $R^2_{quark}$ (in
  $GeV^{-2}$)
  for  $\alpha'_P = 0 $(Fig.5-b) .The straight
  line  shows the experimental value of $\sigma_{eff}$. }
\label{rat11}
\end{figure}

We calculate the case of symmetrical pair production, where  $m=1$
and $y_1=y_2=\frac{Y}{2}$, and we take a ``hard'' Pomeron
in this calculation, $\alpha'_P=0$. We have already considered the
important question of the numerical value of the triple Pomeron vertex
$\gamma$. Fitting the diffraction dissociation data we obtained that
$\frac{\gamma}{g_0}\,\sim\,0.014\,$. This is not such a small number, the
one loop corrections influence and change our results. Indeed, in
Fig.~\ref{rat11} the value of $\sigma_{eff}$ is plotted as a function
$R^2_{quark}$, where we used Eq.~\ref{SIEFF1} for the calculations.

We see that indeed, the corrections change the result, obtained in the
previous calculation. However, the change is not so crucial.
For a ``hard'' Pomeron the value of $\sigma_{eff}\,\approx\,15 mb$
is achieved for $R_{quark}^2\,\approx\,0.08\,-\,0.07\,\, GeV^{-2}$.
The conclusion is that in order to explain the CDF result for the double
parton cross section we need the triple Pomeron vertex, but the value
of $R_{quark}^2$ still remains very small.

\subsection{R(t) in one loop calculation}

\begin{figure}[hptb]
\begin{center}
\psfig{file=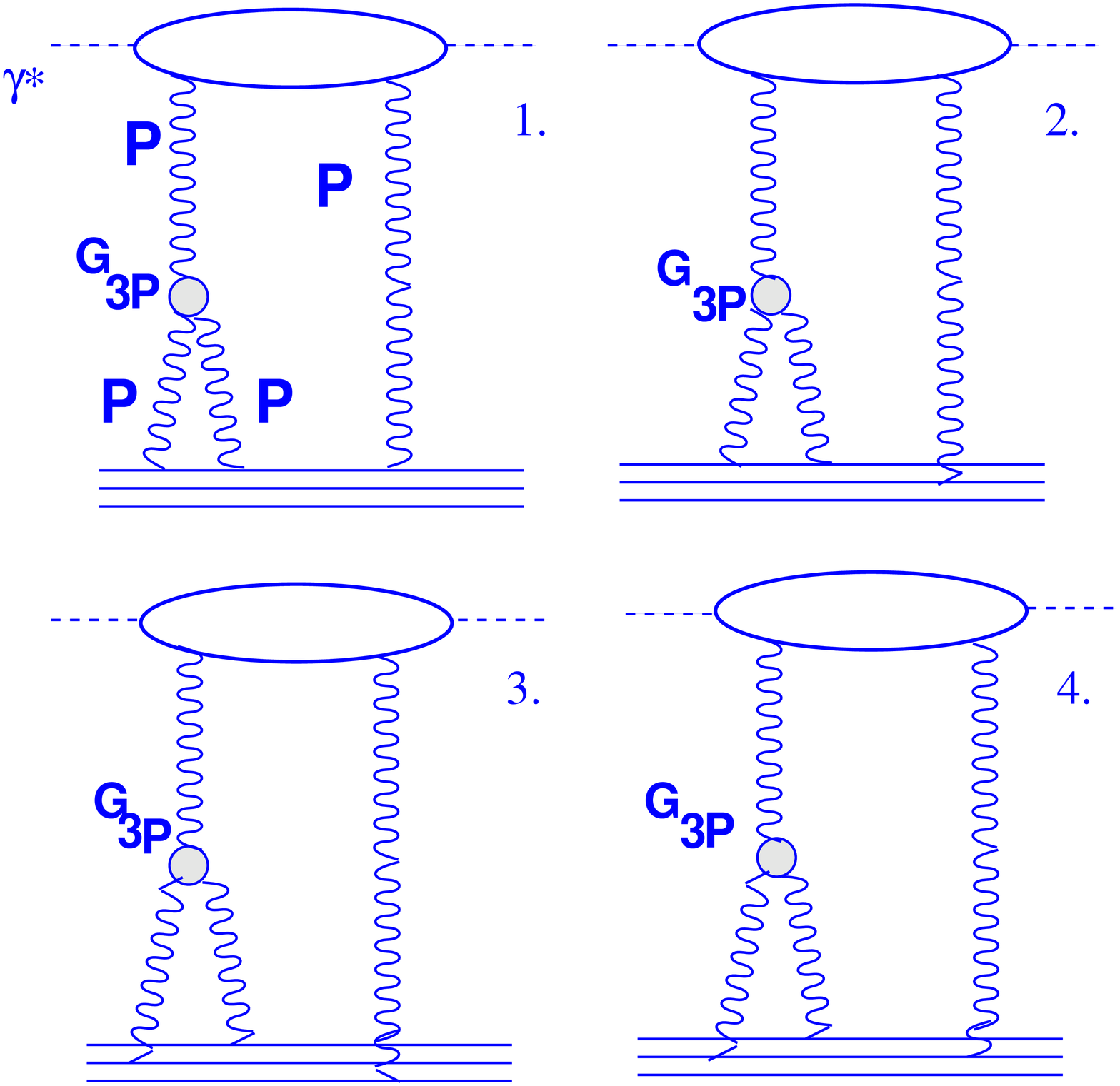, width=80mm}
\end{center}
\caption{\it The $\gamma$ corrections to the inelastic diffraction in DIS . }
\label{Dip4}
\end{figure}

\begin{figure}[hptb]
\begin{center}
\psfig{file=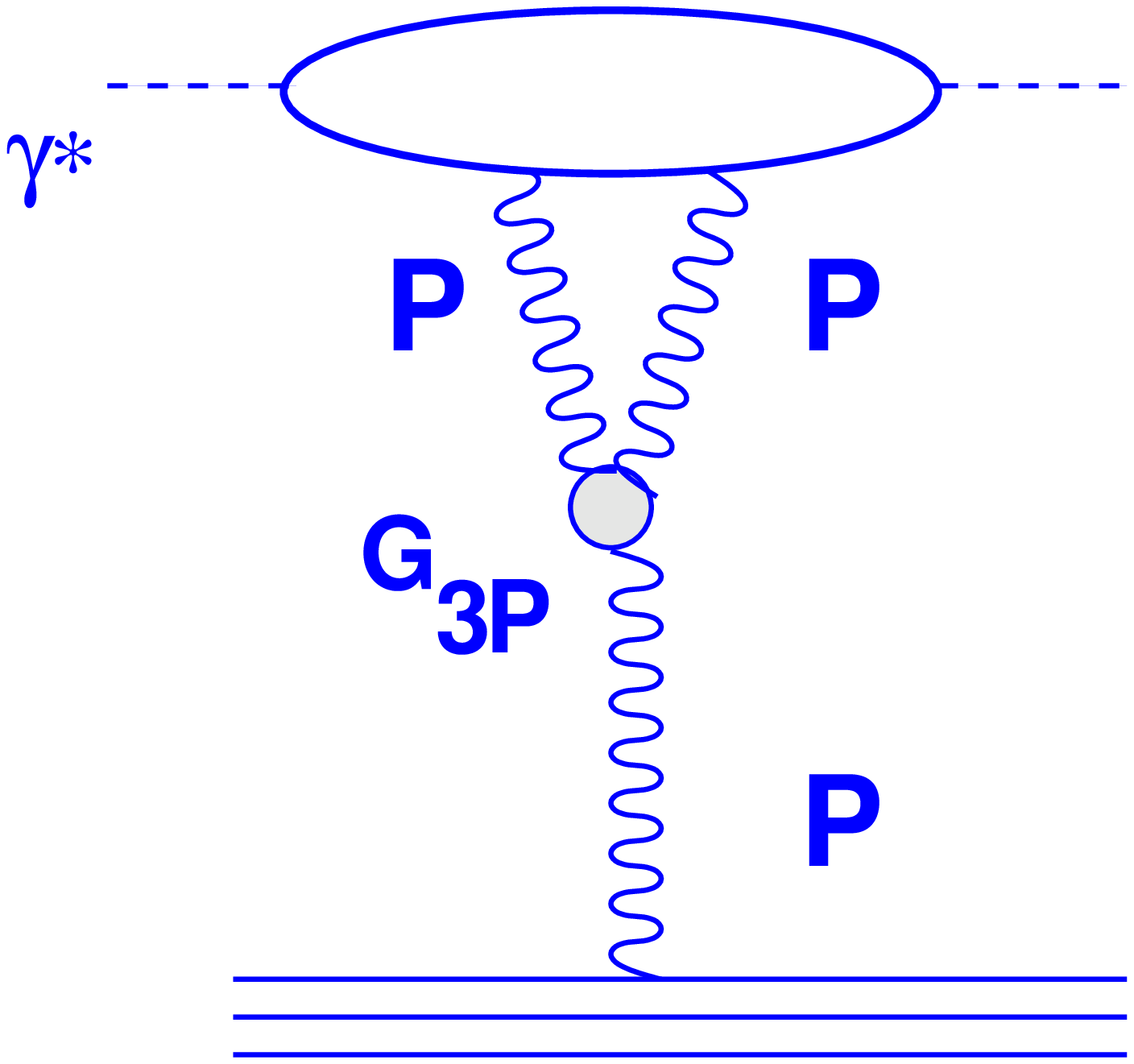, width=80mm}
\end{center}
\caption{\it Inelastic diffraction diagram in DIS . }
\label{Dip44}
\end{figure}

In the case of triple Pomeron interaction we have an additional number of
diagrams which contribute to the elastic cross section of DIS and to the
inelastic diffraction in DIS. First of all, consider the diagrams of DIS,
presented in Figs.~\ref{Dip4} and ~\ref{Dip44}, which describe elastic and
proton diffraction processes together. The expressions corresponding to
these diagrams 
are also presented in the Appendix.

In these diagrams we defined by $\Delta_S\,\approx\,0.08$ and
$\alpha'_P\,\approx\,0.2$ the intercept and the
slope of the ``soft'' Pomeron, and by $\Delta\,\approx\,0.3$
and $\alpha'_H\,\approx\,0$  the intercept and the
slope of the ``hard'' Pomeron. We also introduced
$\tilde{\Delta}\,=\,2\,\Delta_S\,-\,\Delta$.

\begin{figure}[hptb]
\begin{tabular}{ c c}
\psfig{file=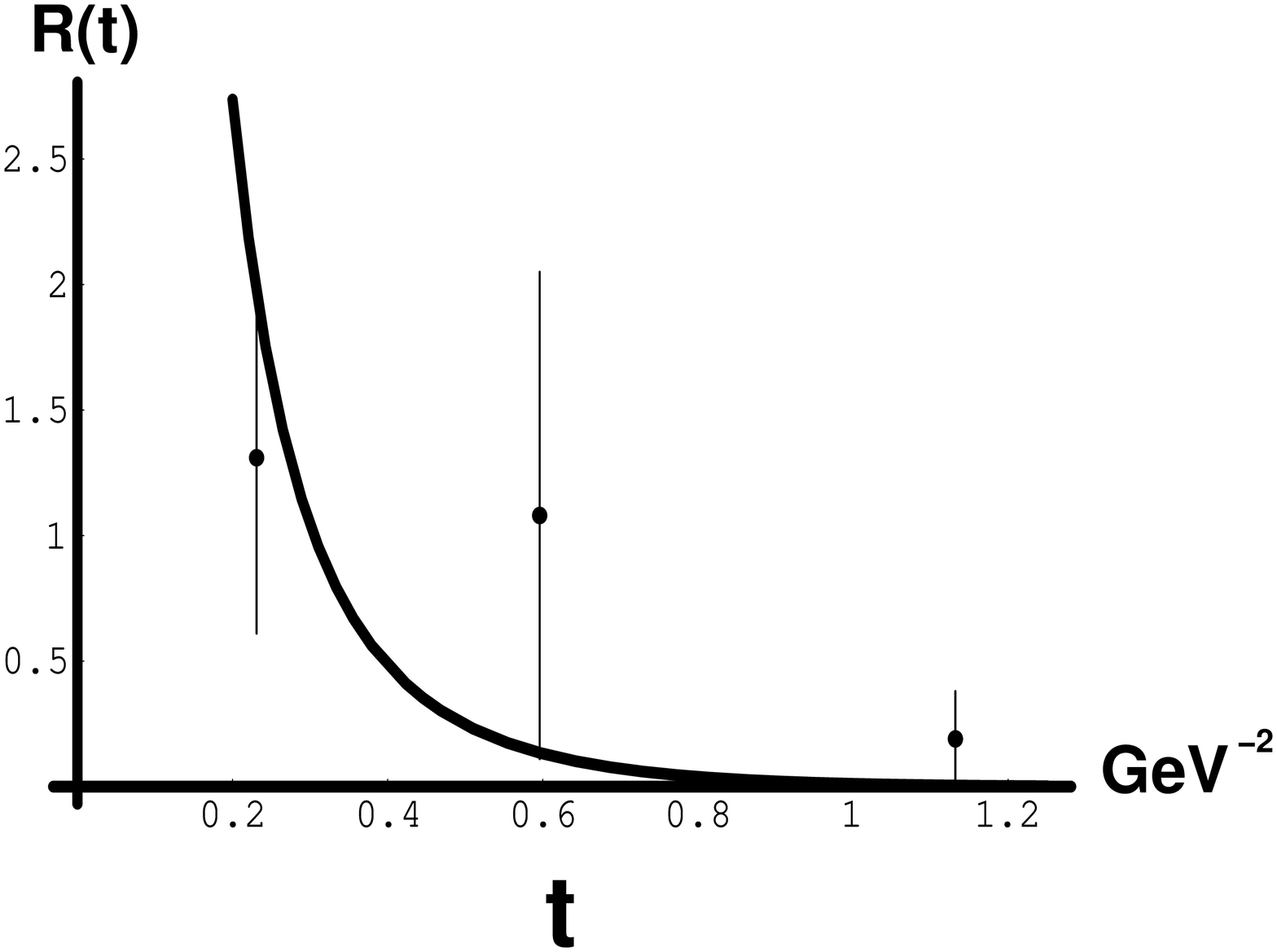,width=80mm} &
\psfig{file=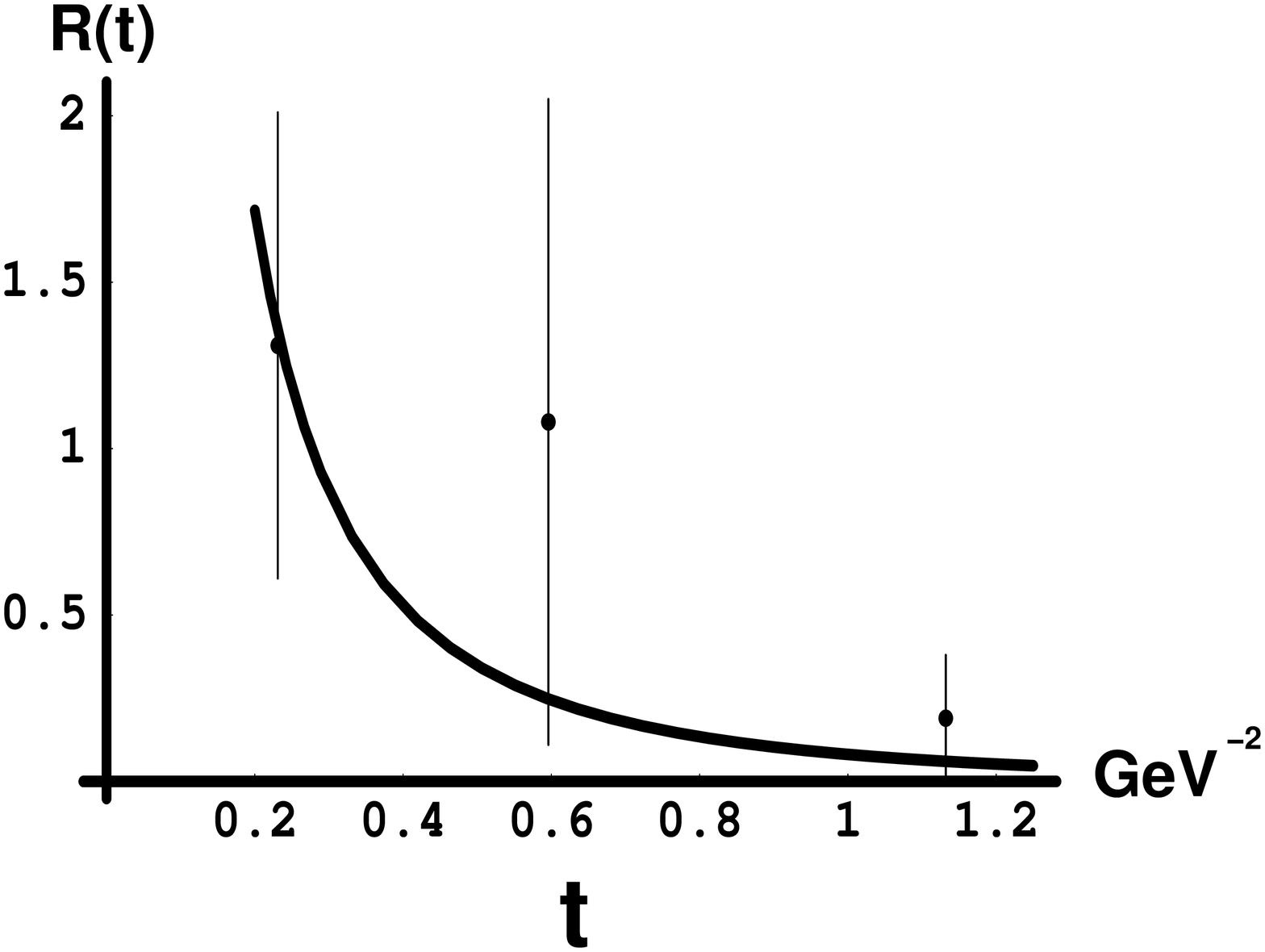,width=80mm}\\
       &  \\
Fig.19-a & Fig. 19-b \\
 &  \\
\end{tabular}
\caption{\it $R(t)$ versus t in one loop calculation for the case of our
parametrization of the Pomeron-quark interaction vertex, Fig.19-a and for
the case of a more realistic parameterization, Fig.19-b.}
\label{rat22}
\end{figure}

The calculation of the  elastic cross section in DIS for one loop
leads to:

$$
\frac{d\sigma^{el}_2}{d t}\,\,=-
\,\,\frac{9}{8\,\pi}
g^4_0\,e^{2\,\Delta\,Y}\,\Le\frac{\gamma}{g_0}\Ra\,
\int_0^{Y}\,\frac{dx\,e^{\tilde{\Delta}\,x}}{\alpha'_P\,x\,+
\,\frac{1}{4\,\alpha}}
\,e^{- |t|(2\,\alpha'_H\,Y + R^2_{quark}
-\,\alpha'_H\,x\,+\frac{\alpha'_P\,x}{2}+\frac{7}{24\,\alpha})}\,-
$$
\beq
-\frac{9}{16\,\pi}
g^4_0\,e^{2\,\Delta\,Y}\,\Le\frac{\gamma}{g_0}\Ra\,
\int_0^{Y}\,\frac{dx\,e^{\tilde{\Delta}\,x}}{\alpha'_P\,x\,+\,R_{quark}^2}
\,e^{- |t|(2\,\alpha'_H\,Y +\frac{3\,R^2_{quark}}{2}
-\,\alpha'_H\,x\,+\frac{\alpha'_P\,x}{2}+\frac{1}{6\,\alpha})}\,\,.
\eeq
Here $\frac{d\sigma^{DIS}_{inel}}{dt}$ is considered as the difference
between the contributions of diagrams of Figs.~\ref{Dip4}, ~\ref{Dip44}
and of the elastic cross section. Now we are ready get the ratio $R(t)$,
which is given by \eq{RA}. The result is shown in Fig.~\ref{rat22}-a.
Actually, there is a fast exponential fall of the curve, due the
parametrization chosen for our vertices. Therefore we present also
the graph Fig.~\ref{rat22}-b, where, instead of the simplest Gaussian
parametrization $e^{t\,R^2_{quark}}$, we introduce a more realistic form
factor $F_{elastic}(|t|)=1/(1\,+\,|t|\,/\,0.72\,)^2$ for the calculation
of the elastic cross section. For the calculation of the diffraction
dissociation cross section we also take a different form factor
$F_{diffr}(|t|)=1/(1\,+\,|t|/2\alpha)$, which is the first term of the
expansion of $e^{-|t|/2\alpha}$, which we used before as the
parametrization for the form factor in the case of diffractive
dissotiacion at the ``tree'' level.

We see, that we obtain here a better fit.

The obtained result for the one loop calculation is, in this simplest
estimations, not so different from the ``tree'' calculation, in spite of
the fact that in one loop calculation the radius of the quark,
$R^2_{quark}\approx\,0.08\,-\,0.1\,\, GeV^{-2}$ is involved.
And it seems that in the order to obtain a better fit to this data,
we have to consider corrections, where more realistic form factors
will be taken into account.

\section{Conclusion}

In this paper we demonstrate that the new beautiful experimental
data on double parton shower effective cross section (CDF, Tevatron)
and on the ratio of elastic and inelastic diffraction production in DIS
(ZEUS, HERA) can be described in the framework of the naive Constituent
Quark Model (CQM). The value of the quark radius turns out to be small:
it is equal to $R^{2}_{quark}\,=\,0.07-\,0.1\,GeV^{-2}$.
This smallness can be considered as an argument supporting the idea
that constituent quarks are the correct degrees of freedom for soft
(long distance) interactions. It should be stressed also that we reached
a satisfactory description of the experimental data by introducing only
the triple Pomeron interaction. This means that the constituent quarks
do not exhaust all degrees of freedom of soft high energy interaction.
Much work is needed to build a comprehensive theoretical approach
for long distance interaction and we consider the fact that the CQM
describes all data of soft interaction, as our small contribution
to the solution of this complicated problem.    

\section{Acknowledgments}
We would like to thank Asher Gotsman, Uri Maor and Eran Naftali
for many informative and encouraging discussions on the sabject of 
this paper.
 The research of S.B. and E.L.  was supported in part by the  GIF 
grant  
\# I-620-22.14/1994, by the BSF grant \# 9800276 and by  the Israel 
Science
Foundation founded by the Israeli Academy of Science
and Humanities.

\newpage
{ \bf \Huge Appendix}
\appendix
\section{The next order corrections}

The next order corrections are the corrections where
the double Pomeron exchange and triple Pomeron vertex
are included. In the case of the total cross section
there are the following additional contributions:

$$
\sigma_{total}^{1}\,\Le\,Y\,\Ra=-
\frac{9}{32\,\pi}\,\frac{g_0^4\,e^{2\,\Delta\,Y}}
{\alpha'_P\,Y\,+\,2\,R^2_{quark}}\,-\,
\frac{9}{4\,\pi}\,\frac{g_0^4\,\alpha\,e^{2\,\Delta\,Y}}
{1+2\,\alpha\,\alpha'_P\,Y\,}\,-\,
\frac{9}{2\,\pi}\,\frac{g_0^4\,\alpha\,e^{2\,\Delta\,Y}}
{1+4\,\alpha\,(\alpha'_P\,Y\,+\,R^2_{quark})}\,-\,
$$
$$
\,\frac{9}{8\,\pi}\,
\,g_0^4\,e^{\Delta\, Y}\,\Le\frac{\gamma}{g_0}\Ra\,
\int_{0}^{Y}\,\frac{dx\,e^{\Delta\,x}}{R_{quark}^2+\alpha'_P\, x}
-\,\frac{9}{\pi}\,
\,g_0^4\,e^{\Delta\, Y}\,\alpha\,\Le\frac{\gamma}{g_0}\Ra\,
\int_{0}^{Y/2}\,\frac{dx\,e^{\Delta\,x}}{1+4\,\alpha\,\alpha'_P\, x}\,\,+
$$
$$
\,\frac{9}{128\,\pi^2}\,
\,g_0^6\,e^{2\,\Delta\, Y}\,\Le\frac{\gamma}{g_0}\Ra\,
\int_{0}^{Y}\,\frac{dx\,e^{\Delta\,x}}
{\Le 2\,\alpha'\,Y+\frac{7\,R_{quark}^2}{2}-\frac{\alpha'_P\,x}{2}\Ra\,
\Le R_{quark}^2+\alpha'_P\,x\Ra}\,+
$$
$$
\,\frac{9}{64\,\pi^2}\,
\,g_0^6\,e^{2\,\Delta\, Y}\,\Le\frac{\gamma}{g_0}\Ra\,
\int_{0}^{Y}\,\frac{dx\,e^{\Delta\,x}}
{\Le 2\,\alpha'\,Y+\frac{7\,R_{quark}^2}{2}-\frac{\alpha'_P\,x}{2}+
\frac{1}{2\,\alpha}\Ra\,\Le R_{quark}^2+\alpha'_P\,x\Ra}\,+
$$
$$
\,\frac{9}{32\,\pi^2}\,
\,g_0^6\,e^{2\,\Delta\, Y}\,\Le\frac{\gamma}{g_0}\Ra\,
\int_{0}^{Y}\,\frac{dx\,e^{\Delta\,x}}
{\Le 4\,\alpha'\,Y+3\,R_{quark}^2-\alpha'_P\,x+
\frac{1}{\alpha}\Ra\,\Le R_{quark}^2+\alpha'_P\,x\Ra}\,+
$$
$$
\,\frac{9}{16\,\pi^2}\,
\,g_0^6\,e^{2\,\Delta\, Y}\,\Le\frac{\gamma}{g_0}\Ra\,
\int_{0}^{Y}\,\frac{dx\,e^{\Delta\,x}}
{\Le 4\,\alpha'\,Y+3\,R_{quark}^2-\alpha'_P\,x+
\frac{3}{2\,\alpha}\Ra\,\Le R_{quark}^2+\alpha'_P\,x\Ra}\,+
$$
$$
\,\frac{9}{64\,\pi^2}\,
\,g_0^6\,e^{2\,\Delta\, Y}\,\Le\frac{\gamma}{g_0}\Ra\,
\int_{0}^{Y}\,\frac{dx\,e^{\Delta\,x}}
{\Le\Le 3\,R_{quark}^2+2\,\alpha'_P\,Y\Ra
\Le\frac{1}{4\,\alpha}+\alpha'_P\,x\Ra-
\frac{\alpha^{'2}_P\,x^2}{2}\Ra}\,-
$$
$$
\,\frac{9}{64\,\pi^2}\,
\,g_0^6\,e^{2\,\Delta\, Y}\,\Le\frac{\gamma}{g_0}\Ra\,
\int_{0}^{Y}\,\frac{dx\,e^{\Delta\,x}}
{\Le 2\,\alpha'\,Y+2\,R_{quark}^2-\frac{\alpha'_P\,x}{2}+
\frac{3}{8\,\alpha}\Ra\,
\Le\frac{1}{4\,\alpha}+\alpha'_P\,x\Ra}\,+
$$
$$
\,\frac{9}{32\,\pi^2}\,
\,g_0^6\,e^{2\,\Delta\, Y}\,\Le\frac{\gamma}{g_0}\Ra\,
\int_{0}^{Y}\,\frac{dx\,e^{\Delta\,x}}
{\Le 2\,\alpha'_P\,Y-\frac{\alpha'_P\,x}{2}+
\frac{7}{8\,\alpha}\Ra\,
\Le\frac{1}{4\,\alpha}+\alpha'_P\,x\Ra}\,+
$$
\beq
\,\frac{9}{32\,\pi^2}\,
\,g_0^6\,e^{2\,\Delta\, Y}\,\Le\frac{\gamma}{g_0}\Ra\,
\int_{0}^{Y}\,\frac{dx\,e^{\Delta\,x}}
{\Le\Le R_{quark}^2+2\,\alpha'_P\,Y+\frac{1}{2\,\alpha}\Ra
\Le\frac{1}{4\,\alpha}+\alpha'_P\,x\Ra-
\frac{\alpha^{'2}_P\,x^2}{2}\Ra}\,.
\eeq
For the elastic cross section we have in the next order:
$$
\sigma_{Elastic}^{1}\,\Le\,Y\,\Ra=-
\,\frac{81}{128\,\pi^2}\,
\,g_0^6\,e^{2\,\Delta\, Y}\,\Le\frac{\gamma}{g_0}\Ra\,
\int_{0}^{Y}\,\frac{dx\,e^{\Delta\,x}}
{\Le 2\,\alpha'\,Y+\frac{3\,R_{quark}^2}{2}-\frac{\alpha'_P\,x}{2}+
\frac{1}{6\,\alpha}\Ra\,
\Le R_{quark}^2+\alpha'_P\,x\Ra}\,-
$$
\beq
\,\frac{81}{64\,\pi^2}\,
\,g_0^6\,e^{2\,\Delta\, Y}\,\Le\frac{\gamma}{g_0}\Ra\,
\int_{0}^{Y}\,\frac{dx\,e^{\Delta\,x}}
{\Le 2\,\alpha'\,Y+R_{quark}^2-\frac{\alpha'_P\,x}{2}+
\frac{13}{24\,\alpha}\Ra\,\Le \frac{1}{4\,\alpha}+\alpha'_P\,x\Ra}\,.
\eeq
And for sum of elastic and diffractive dissotiation cross sections
we have:
$$
\sigma_{Elastic+diffr}^{1}\,\Le\,Y\,\Ra=
\,\frac{9}{8\,\pi}\,
\,g_0^4\,e^{\Delta\, Y}\,\Le\frac{\gamma}{g_0}\Ra\,
\int_{0}^{Y}\,\frac{dx\,e^{\Delta\,x}}{R_{quark}^2+\alpha'_P\, x}.
+\,\frac{9}{\pi}\,
\,g_0^4\,e^{\Delta\, Y}\,\alpha\,\Le\frac{\gamma}{g_0}\Ra\,
\int_{0}^{Y/2}\,\frac{dx\,e^{\Delta\,x}}{1+4\,\alpha\,\alpha'_P\, x}\,\,-
$$
$$
\,\frac{27}{128\,\pi^2}\,
\,g_0^6\,e^{2\,\Delta\, Y}\,\Le\frac{\gamma}{g_0}\Ra\,
\int_{0}^{Y}\,\frac{dx\,e^{\Delta\,x}}
{\Le 2\,\alpha'\,Y+\frac{7\,R_{quark}^2}{2}-\frac{\alpha'_P\,x}{2}\Ra\,
\Le R_{quark}^2+\alpha'_P\,x\Ra}\,-
$$
$$
\,\frac{27}{64\,\pi^2}\,
\,g_0^6\,e^{2\,\Delta\, Y}\,\Le\frac{\gamma}{g_0}\Ra\,
\int_{0}^{Y}\,\frac{dx\,e^{\Delta\,x}}
{\Le 2\,\alpha'\,Y+\frac{7\,R_{quark}^2}{2}-\frac{\alpha'_P\,x}{2}+
\frac{1}{2\,\alpha}\Ra\,\Le R_{quark}^2+\alpha'_P\,x\Ra}\,-
$$
$$
\,\frac{27}{32\,\pi^2}\,
\,g_0^6\,e^{2\,\Delta\, Y}\,\Le\frac{\gamma}{g_0}\Ra\,
\int_{0}^{Y}\,\frac{dx\,e^{\Delta\,x}}
{\Le 4\,\alpha'\,Y+3\,R_{quark}^2-\alpha'_P\,x+
\frac{1}{\alpha}\Ra\,\Le R_{quark}^2+\alpha'_P\,x\Ra}\,-
$$
$$
\,\frac{27}{16\,\pi^2}\,
\,g_0^6\,e^{2\,\Delta\, Y}\,\Le\frac{\gamma}{g_0}\Ra\,
\int_{0}^{Y}\,\frac{dx\,e^{\Delta\,x}}
{\Le 4\,\alpha'\,Y+3\,R_{quark}^2-\alpha'_P\,x+
\frac{3}{2\,\alpha}\Ra\,\Le R_{quark}^2+\alpha'_P\,x\Ra}\,-
$$
$$
\,\frac{27}{64\,\pi^2}\,
\,g_0^6\,e^{2\,\Delta\, Y}\,\Le\frac{\gamma}{g_0}\Ra\,
\int_{0}^{Y}\,\frac{dx\,e^{\Delta\,x}}
{\Le\Le 3\,R_{quark}^2+2\,\alpha'_P\,Y\Ra
\Le\frac{1}{4\,\alpha}+\alpha'_P\,x\Ra-
\frac{\alpha^{'2}_P\,x^2}{2}\Ra}\,-
$$
$$
\,\frac{27}{64\,\pi^2}\,
\,g_0^6\,e^{2\,\Delta\, Y}\,\Le\frac{\gamma}{g_0}\Ra\,
\int_{0}^{Y}\,\frac{dx\,e^{\Delta\,x}}
{\Le 2\,\alpha'\,Y+2\,R_{quark}^2-\frac{\alpha'_P\,x}{2}+
\frac{3}{8\,\alpha}\Ra\,
\Le\frac{1}{4\,\alpha}+\alpha'_P\,x\Ra}\,-
$$
$$
\,\frac{27}{32\,\pi^2}\,
\,g_0^6\,e^{2\,\Delta\, Y}\,\Le\frac{\gamma}{g_0}\Ra\,
\int_{0}^{Y}\,\frac{dx\,e^{\Delta\,x}}
{\Le 2\,\alpha'_P\,Y-\frac{\alpha'_P\,x}{2}+
\frac{7}{8\,\alpha}\Ra\,
\Le\frac{1}{4\,\alpha}+\alpha'_P\,x\Ra}\,-
$$
\beq
\,\frac{27}{32\,\pi^2}\,
\,g_0^6\,e^{2\,\Delta\, Y}\,\Le\frac{\gamma}{g_0}\Ra\,
\int_{0}^{Y}\,\frac{dx\,e^{\Delta\,x}}
{\Le\Le R_{quark}^2+2\,\alpha'_P\,Y+\frac{1}{2\,\alpha}\Ra
\Le\frac{1}{4\,\alpha}+\alpha'_P\,x\Ra-
\frac{\alpha^{'2}_P\,x^2}{2}\Ra}\,.
\eeq

\section{Diagrams which contribute to the double inclusive
cross section}

\begin{enumerate}

\item First diagram of Fig.~\ref{doub1}:

\beq
D_{4}(Y)\,=\,\frac{9}{4\,\pi}\,
\,g_0^4\,e^{\Delta\, Y}\,\Le\frac{\gamma}{g_0}\Ra\,
\int^{Y}_{Y/2}\,\frac{dx\,e^{\Delta\,x}}{R_{quark}^2+\alpha'_P\, x}.
\eeq

\item Second diagram of Fig.~\ref{doub1}:

\beq
D_{5}(Y)\,=\,\frac{18}{\pi}\,
\,g_0^4\,e^{\Delta\, Y}\,\alpha\,\Le\frac{\gamma}{g_0}\Ra\,
\int^{Y}_{Y/2}\,\frac{dx\,e^{\Delta\,x}}{1+4\alpha\,\alpha'_P\, x}.
\eeq

\item First diagram of Fig.~\ref{pp3P}

\beq
D_{6}(Y)\,=-\,\frac{9}{64\,\pi^2}\,
\,g_0^6\,e^{2\,\Delta\, Y}\,\Le\frac{\gamma}{g_0}\Ra\,
\int_{0}^{Y/2}\,\frac{dx\,e^{\Delta\,x}}
{\Le 2\,\alpha'\,Y+\frac{7\,R_{quark}^2}{2}-\frac{\alpha'_P\,x}{2}\Ra\,
\Le R_{quark}^2+\alpha'_P\,x\Ra}.
\eeq

\item Second diagram of Fig.~\ref{pp3P} gives

\beq
D_{7}(y)\,=-\,\frac{9}{32\,\pi^2}\,
\,g_0^6\,e^{2\,\Delta\, Y}\,\Le\frac{\gamma}{g_0}\Ra\,
\int_{0}^{Y/2}\,\frac{dx\,e^{\Delta\,x}}
{\Le 2\,\alpha'\,Y+\frac{7\,R_{quark}^2}{2}-\frac{\alpha'_P\,x}{2}+
\frac{1}{2\,\alpha}\Ra\,\Le R_{quark}^2+\alpha'_P\,x\Ra}.
\eeq

\item Third diagram of Fig.~\ref{pp3P} gives

\beq
D_{8}(Y)\,=-\,\frac{9}{16\,\pi^2}\,
\,g_0^6\,e^{2\,\Delta\, Y}\,\Le\frac{\gamma}{g_0}\Ra\,
\int_{0}^{Y/2}\,\frac{dx\,e^{\Delta\,x}}
{\Le 4\,\alpha'\,Y+3\,R_{quark}^2-\alpha'_P\,x+
\frac{1}{\alpha}\Ra\,\Le R_{quark}^2+\alpha'_P\,x\Ra}.
\eeq

\item Fourth diagram of Fig.~\ref{pp3P}

\beq
D_{9}(Y)\,=-\,\frac{9}{8\,\pi^2}\,
\,g_0^6\,e^{2\,\Delta\, Y}\,\Le\frac{\gamma}{g_0}\Ra\,
\int_{0}^{Y/2}\,\frac{dx\,e^{\Delta\,x}}
{\Le 4\,\alpha'\,Y+3\,R_{quark}^2-\alpha'_P\,x+
\frac{3}{2\,\alpha}\Ra\,\Le R_{quark}^2+\alpha'_P\,x\Ra}.
\eeq

\item Fifth diagram of  Fig.~\ref{pp3P}

\beq
D_{10}(Y)\,=-\,\frac{9}{32\,\pi^2}\,
\,g_0^6\,e^{2\,\Delta\, Y}\,\Le\frac{\gamma}{g_0}\Ra\,
\int_{0}^{Y/2}\,\frac{dx\,e^{\Delta\,x}}
{\Le\Le 3\,R_{quark}^2+2\,\alpha'_P\,Y\Ra
\Le\frac{1}{4\,\alpha}+\alpha'_P\,x\Ra-
\frac{\alpha^{'2}_P\,x^2}{2}\Ra}.
\eeq

\item Sixth diagram of Fig.~\ref{pp3P}

\beq
D_{11}(Y)\,=-\,\frac{9}{32\,\pi^2}\,
\,g_0^6\,e^{2\,\Delta\, Y}\,\Le\frac{\gamma}{g_0}\Ra\,
\int_{0}^{Y/2}\,\frac{dx\,e^{\Delta\,x}}
{\Le 2\,\alpha'\,Y+2\,R_{quark}^2-\frac{\alpha'_P\,x}{2}+
\frac{3}{8\,\alpha}\Ra\,
\Le\frac{1}{4\,\alpha}+\alpha'_P\,x\Ra}.
\eeq

\item Seventh diagram of Fig.~\ref{pp3P}

\beq
D_{12}(Y)\,=-\,\frac{9}{16\,\pi^2}\,
\,g_0^6\,e^{2\,\Delta\, Y}\,\Le\frac{\gamma}{g_0}\Ra\,
\int_{0}^{Y/2}\,\frac{dx\,e^{\Delta\,x}}
{\Le 2\,\alpha'_P\,Y-\frac{\alpha'_P\,x}{2}+
\frac{7}{8\,\alpha}\Ra\,
\Le\frac{1}{4\,\alpha}+\alpha'_P\,x\Ra}.
\eeq

\item Eighth  diagram of Fig.~\ref{pp3P}

\beq
D_{13}(Y)\,=-\,\frac{9}{16\,\pi^2}\,
\,g_0^6\,e^{2\,\Delta\, Y}\,\Le\frac{\gamma}{g_0}\Ra\,
\int_{0}^{Y/2}\,\frac{dx\,e^{\Delta\,x}}
{\Le\Le R_{quark}^2+2\,\alpha'_P\,Y+\frac{1}{2\,\alpha}\Ra
\Le\frac{1}{4\,\alpha}+\alpha'_P\,x\Ra-
\frac{\alpha^{'2}_P\,x^2}{2}\Ra}.
\eeq

\end{enumerate}

\section{Diagrams wich contribute to the total inelastic
diffraction in DIS}
\begin{enumerate}

\item The first diagram of Fig.~\ref{Dip4} gives

\beq \label{DISP11}
\frac{d\sigma^D_3}{d t}\,\,=-
\,\,\frac{3}{16\,\pi}
g^4_0\,e^{2\,\Delta\,Y}\,\Le\frac{\gamma}{g_0}\Ra\,
\int_0^{Y}\,\frac{dx\,e^{\tilde{\Delta}\,x}}
{\alpha'_P\,x\,+\,R_{quark}^2}
\,e^{- |t|(2\, \alpha'_H\,Y +\frac{3\,R^2_{quark}}{2}
-\,\alpha'_H\,x\,+\frac{\alpha'_P\,x}{2})}\,\,.
\eeq

\item From second diagram of Fig.~\ref{Dip4} we obtain

\beq \label{DISP22}
\frac{d\sigma^D_4}{d t}\,\,=-
\,\,\frac{3}{8\,\pi}
g^4_0\,e^{2\,\Delta\,Y}\,\Le\frac{\gamma}{g_0}\Ra\,
\int_0^{Y}\,\frac{dx\,e^{\tilde{\Delta}\,x}}{\alpha'_P\,x\,+\,R_{quark}^2}
\,e^{- |t|(2\, \alpha'_H\,Y +\frac{3\,R^2_{quark}}{2}
-\,\alpha'_H\,x\,+\frac{\alpha'_P\,x}{2}+\frac{1}{2\,\alpha})}\,\,.
\eeq

\item Third diagram of Fig.~\ref{Dip4}

\beq \label{DISP33}
\frac{d\sigma^D_5}{d t}\,\,=-
\,\,\frac{3}{8\,\pi}
g^4_0\,e^{2\,\Delta\,Y}\,\Le\frac{\gamma}{g_0}\Ra\,
\int_0^{Y}\,\frac{dx\,e^{\tilde{\Delta}\,x}}{\alpha'_P\,x\,+
\,\frac{1}{4\,\alpha}}
\,e^{- |t|(2\,\alpha'_H\,Y + R^2_{quark}
-\,\alpha'_H\,x\,+\frac{\alpha'_P\,x}{2}+\frac{3}{8\,\alpha})}\,\,.
\eeq

\item Fourth diagram of Fig.~\ref{Dip4} gives

\beq \label{DISP44}
\frac{d\sigma^D_6}{d t}\,\,=-
\,\,\frac{3}{8\,\pi}
g^4_0\,e^{2\,\Delta\,Y}\,\Le\frac{\gamma}{g_0}\Ra\,
\int_0^{Y}\,\frac{dx\,e^{\tilde{\Delta}\,x}}{\alpha'_P\,x\,+
\,\frac{1}{4\,\alpha}}
\,e^{- |t|(2\,\alpha'_H\,Y + R^2_{quark}
-\,\alpha'_H\,x\,+\frac{\alpha'_P\,x}{2}+\frac{1}{8\,\alpha}-
\frac{1}{8\,\alpha+32\,\alpha^2\,\alpha'_P\,x})}\,\,,
\eeq

\item And the last diagram of Fig.~\ref{Dip44}
leads to:

\beq\label{DIS55}
\frac{d\sigma^D_7}{d t}\,\,=
6\,g^2_0\,\Le\frac{\gamma}{g_0}\Ra\,e^{- |t|\, R^2_{quark}}\,
\frac{1\,-\,e^{-(2\Delta-\Delta_S)\,Y}}{2\Delta-\Delta_S}\,\,.
\eeq

\end{enumerate}

\end{document}